\documentclass[aps,
pre,
twocolumn,
longbibliography,
superscriptaddress]{revtex4-1}

\usepackage{graphicx}
\usepackage{amssymb,amsmath,physics}
\usepackage{float}
\usepackage{xcolor}
\allowdisplaybreaks

\begin{document}

\renewcommand{\floatpagefraction}{0.9}

\title{Analytical Interaction Potential for Lennard-Jones Rods}

\author{Junwen Wang}
\affiliation{Department of Mechanical Engineering, Virginia Tech, Blacksburg, Virginia 24061, USA}
\affiliation{Center for Soft Matter and Biological Physics, Virginia Tech, Blacksburg, Virginia 24061, USA}
\affiliation{Macromolecules Innovation Institute, Virginia Tech, Blacksburg, Virginia 24061, USA}
\author{Gary Seidel}
\affiliation{Department of Aerospace and Ocean Engineering, Virginia Tech, Blacksburg, Virginia 24061, USA}
\affiliation{Macromolecules Innovation Institute, Virginia Tech, Blacksburg, Virginia 24061, USA}
\affiliation{Department of Mechanical Engineering, Virginia Tech, Blacksburg, Virginia 24061, USA}
\author{Shengfeng Cheng}
\email{chengsf@vt.edu}
\affiliation{Department of Physics, Virginia Tech, Blacksburg, Virginia 24061, USA}
\affiliation{Center for Soft Matter and Biological Physics, Virginia Tech, Blacksburg, Virginia 24061, USA}
\affiliation{Macromolecules Innovation Institute, Virginia Tech, Blacksburg, Virginia 24061, USA}
\affiliation{Department of Mechanical Engineering, Virginia Tech, Blacksburg, Virginia 24061, USA}

\begin{abstract}
An analytical form has been derived using Ostrogradski's integration method for the interaction between two thin rods of finite lengths in arbitrary relative configurations in a 3-dimensional space, each treated as a line of material points interacting through the Lennard-Jones 12-6 potential. Simplified analytical forms for coplanar, parallel, and collinear rods are also derived. Exact expressions for the force and torque between the rods are obtained. Similar results for a point particle interacting with a thin rod are provided. These interaction potentials can be widely used for analytical descriptions and computational modeling of systems involving rod-like objects such as liquid crystals, colloids, polymers, elongated viruses and bacteria, and filamentous materials including carbon nanotubes, nanowires, biological filaments, and their bundles.
\end{abstract}

\date{\today}


\maketitle

\section{Introduction}

First proposed to explain the behavior of gases, van der Waals (vdW) forces are ubiquitous among atoms, molecules, clusters, nanostructures, and macrobodies, underscoring their unique role in determining the properties of materials and condensed matter.\cite{Parsegian2005book, Wang2022ACSNano, Fiedler2023} Dispersive vdW forces are fundamentally a Casimir effect, originating from the electrodynamic correlations of charge density fluctuations in interacting bodies.\cite{CasimirPhysicsBook, Brevik1999,Ambrosetti2016} London was the first to show that an attractive potential exists between two neutral atoms, and that is scales as $1/r^6$ in the nonretarded regime with $r$ being the interatomic separation. The system is in this nonretarded regime when $r$ is smaller than the characteristic wavelengths of the atomic absorption spectra.\cite{London1930aEng, London1930bEng, London1930cEng, London1937} The $1/r^6$ vdW-London potential has been directly confirmed in an experiment with Rydberg atoms by B\'{e}guin et al.\cite{Beguin2013} More than a decade after London's seminal work, Casimir and Polder showed via perturbation calculations that in the retarded regime, the distance-dependence of the attractive potential becomes $1/r^7$.\cite{Casimir1948} This result was subsequently confirmed in the calculation of Dzyaloshinskii using the Feynman diagram technique.\cite{Dzyaloshinskii1957}

With the understanding of the atomic origin of vdW-London forces, the focus shifted to computing such forces for ensembles of atoms ranging from molecules to macrobodies. Along the way, various methods have been attempted. Casimir and Polder calculated the interaction between a neutral atom and a perfectly conducting plane using quantum electrodynamics and perturbation methods.\cite{Casimir1948} Casimir further showed that similar expressions can be obtained by using classical electrodynamics to compute the change of zero-point energies and applied the method to calculate the vdw-London attraction between two metal plates, revealing the famous Casimir force.\cite{Casimir1948b} Lifshitz established a macroscopic theory of intermolecular forces between solids by introducing a random field into the Maxwell equations, where many-body and retardation effects are naturally captured.\cite{Lifshitz1956} This theory was later reformulated and generalized by Dzyaloshinskii et al. using the language of quantum field theory.\cite{Dzyaloshinskii1961} A simpler macroscopic derivation of Lifshitz's result on the non-retarded vdW interaction between two dielectric media was later provided by van Kampen et al.\cite{vanKampen1968}

Many other theories and methods \cite{Linder1972, Langbein1973inbook, Mahanty1976book, Parsegian2005book} for explaining and computing the vdW forces between molecules and macrobodies have been proposed, including the reaction field method developed by Linder \cite{Linder1960, Linder1962, Linder1964a, Linder1964b, Linder1972}, the general susceptibility theory of McLachlan \cite{McLachlan1963a, McLachlan1963b, McLachlan1963c, McLachlan1964, McLachlan1965}, and the screened field method based on the Drude-Lorentz model.\cite{London1937, Bade1957a, Bade1957b, Bade1958, Renne1967, Nijboer-Renne1968, Renne1970, Langbein1969, Langbein1971b}. In particular, Renne and Nijboer provided an atomistic derivation of Lifshitz's result on the vdW forces between macroscopic bodies by treating atoms as isotropic harmonic oscillators.\cite{Renne1967, Nijboer-Renne1968} Langbein proposed a plane wave method based on the eigenfunctions of the Helmholtz equation to compute the retarded dispersion energy.\cite{Langbein1970} However, it is usually challenging to apply these methods, including the Lifshitz theory, to systems other than half-spaces, plates, and spheres \cite{Kihara1965, Imura1973}. More recently, methods based on the density-functional theory \cite{Dobson2006}, quantum Monte Carlo simulation \cite{Drummond2007, Azadi2018}, and the adiabatic-connection fluctuation-dissipation theorem \cite{Hermann2017} were used to compute the vdW interactions among molecules and condensed bodies including nanostructures.

To deal with complicated shaped bodies, Derjaguin has proposed an approximation method for computing the vdW interaction between macrobodies by summing the potential of parallel facing surface-element pairs, which can be readily approximated as the potential between two plates.\cite{Derjaguin1934} The Derjaguin approximation is valid when the closest-approach distance is much smaller than the sizes of the involved macrobodies but larger than the characteristic length scale of their surface roughness. This method has been extended and improved by others.\cite{White1983, Bhattacharjee1997, Dantchev2012}

In the de Boer-Hamaker approach, the total vdW interaction between two bodies is obtained by summing over all interacting pairs of atoms or molecules which the bodies are composed of.\cite{Bradley1932, deBoer1936, Hamaker1937, Girifalco1956, Salem1962, Israelachvili1973} Obviously, pairwise additivity is assumed in this ``microscopic''approach. For bodies that can be treated as continuous media, the sum becomes a double integral of a pair potential between two volume elements, one on each body.\cite{Vold1954, Clayfield1971, Rosenfield1974, Amadon1991, Gu1999, Tadmor2001, Kirsch2003, He2012, Maeda2015} A standard choice of the pair potential is the Lennard-Jone (LJ) 12-6 potential, which includes not only a $1/r^6$ term describing the nonretarded vdw attraction but also a $1/r^{12}$ term for the Pauli repulsion between atoms.\cite{Lennard-Jones1931} Closed forms of the integrated vdW $1/r^6$ attraction have been reported for various planar and spherical geometries and also for infinitely long cylinders in parallel or perpendicular configurations. These results are nicely summarized in Ref.~\cite{Parsegian2005book}. The integrated form of the full LJ 12-6 potential repulsion can also be obtained for certain geometries and some results have been reported in the literature.\cite{Abraham1977, Abraham1978, Magda1985, Henrard1999, Girifalco2000, Everaers2003, Zhang2004, Sun2006PRB, Vesely2006, Zhbanov2010, Pogorelov2012, Wu2012, Hamady2013, Logan2018}

Understanding the interactions between cylindrical objects is crucial for the description of diverse systems such as liquid crystals \cite{Kats2015}, polymers \cite{vanderSchoot1992, Floyd2021, Lu2015JCP}, colloids \cite{Boles2016}, carbon nanotubes (CNTs) \cite{Siber2009}, nanowires \cite{Ji2012ACSNano, Wang2013APE}, and biological filaments \cite{Cheng2012SM}. Many Viruses and bacteria also have rod-like shapes.\cite{Nijjer2023} It is thus of great interest to integrate the LJ 12-6 potential for two rods. Some attempts have been reported in the literature, mainly in the context of CNT-CNT interactions. Henrard et al. have used the integration approach to evaluate the vdW interaction between CNTs in bundles.\cite{Henrard1999} Girifalco et al. have integrated a LJ carbon-carbon potential for two parallel, infinitely long CNTs.\cite{Girifalco2000} Sun et al. have developed an approximate approach to obtain the vdW potential between two CNTs.\cite{Sun2006PRB} Within the LJ approximation, Popescu et al. have calculated the vdW potential energy between two parallel infinitely long single-walled CNTs that are radially deformed.\cite{Popescu2008} Lu et al. have studied the interaction between walls in multi-walled CNTs by integrating the LJ 12-6 potential.\cite{Lu2009APL} Zhbanov et al. have developed analytical expressions for the vdW potential energy and force between two crossed CNTs.\cite{Zhbanov2010} Pogorelov et al. have reported algebraic expressions for the vdW potentials and forces between two parallel and crossed CNTs.\cite{Pogorelov2012} Vesely has proposed an approximate form of the integrated LJ interaction between two sticks, each made of a homogeneous distribution of LJ centers.\cite{Vesely2006} Recently, Logan and Tkachenko have proposed a compact interaction potential for the vdW attraction of two finite rods at arbitrary angles and separations via interpolating various asymptotic limits.\cite{Logan2018} However, a fully analytical form of the rod-rod interaction based on the LJ potential has been elusive.

In this paper, we present an analytical form of the integrated LJ 12-6 potential between two thin rods at arbitrary relative configurations in 3-dimensions, with the rods having radii that are much smaller than their separation and thus can be approximated as two material lines. By treating each line as a continuous distribution of LJ point particles, we are able to integrate the LJ 12-6 potential between two LJ thin rods with both finite and infinite lengths in arbitrary arrangements. Analytical forms of the forces and torques are derived. Furthermore, analytical results on the integrated potential between a LJ point particle and a LJ thin rod are obtained.

\section{Theoretical Model of Rod-Rod Interactions}

\begin{figure}[htb]
    \centering
    \includegraphics[width=0.45\textwidth]{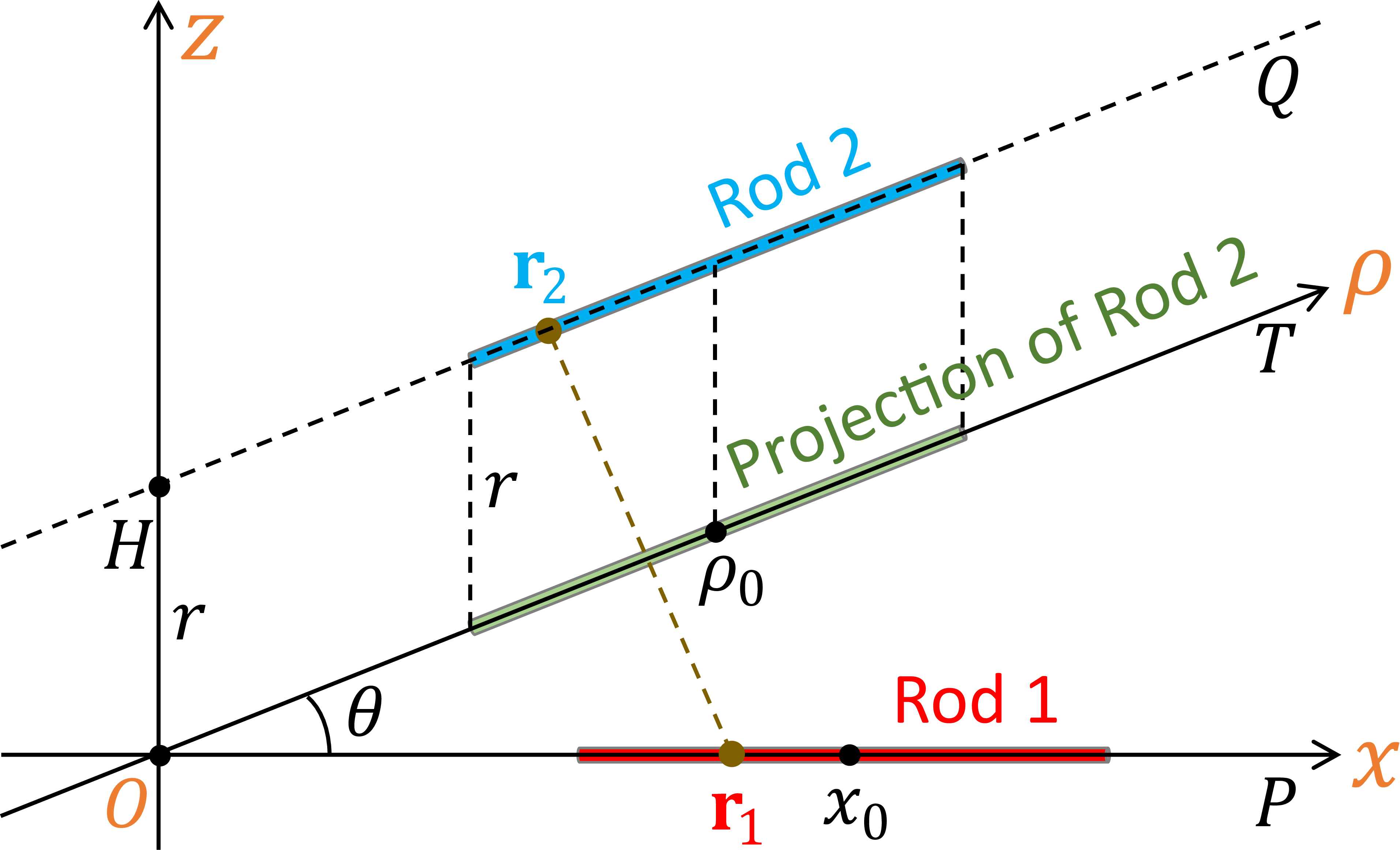}
    \caption{A general skew configuration of two thin rods of finite lengths in 3-dimensions. The $z$-axis is perpendicular to both rod 1 and 2; the $x\rho$-plane is perpendicular to $z$ and contains rod 1 located on the $x$-axis. The projection of rod 2 in this plane is along the $\rho$-axis. The angle between rod 1 and the projection of rod 2 is $\theta$.}
    \label{fg:rod-rod-geo}
\end{figure}

A general skew configuration of two rods in a 3-dimensional space is illustrated in Fig.~\ref{fg:rod-rod-geo}. The axis of rod 1 is on the straight line $OP$ while that of rod 2 is on the straight line $HQ$. The common perpendicular of the two lines is $OH$, and the distance between $O$ and $H$ is denoted as $|r|$, which is the minimal distance between the rod axes. Through point $O$, a straight line $OT$ is drawn to be parallel with $HQ$, and rod 2 is then projected to $OT$ by a translation parallel with $OH$. The angle $\angle POT$ is denoted as $\theta$. A right-handed coordinate system ($x\rho z$) is constructed by defining 3 unit vectors, $\vb{n}_x$, $\vb{n}_{\rho}$, and $\vb{n}_z$, along the rays $OP$, $OT$, and $OH$, respectively. For any relative configurations of two rods, such a coordinate system can always be constructed with $x_0\ge0$, $\rho_0 \ge 0$, $0\le\theta\le\pi$, and $\vb{n}_z = \frac{1}{\sin\theta} \vb{n}_x \times \vb{n}_\rho$. Note that this coordinate frame, with point $O$ as the origin, might not be orthogonal. In this frame, the center of rod 1 is located at $(x_0, 0, 0)$ while the center of the projection of rod 2 is located at $(0, \rho_0, 0)$ and the center of rod 2 is located at $(0, \rho_0, r)$. Using this frame, the configuration of the two rods is completely determined by $x_0$, $\rho_0$, $r$, and $\theta$. Furthermore, in this setting, $r$ is allowed to be negative if rod 2 is below the $x\rho$-plane, but the integrated potentials will be even functions of $r$, as expected.

We will use the following conventions for special configurations. For two parallel rods, $\theta = 0 $ and $|r|$ is the shortest distance between their axes. For two coplanar but nonparallel rods, $r=0$. For collinear rods, we will adopt $r=0$ and $\theta = 0$.

The Lennard-Jones (LJ) potential between two point particles separated at a distance $R$ can be written as
\begin{equation}
\label{eq:lj_potential}
    U(R) = -\frac{A}{R^6} + \frac{B}{R^{12}}~,
\end{equation}
where $A$ and $B$ are two constants governing the interaction strength and characteristic size of the particles.

\subsection{Attractive Potential}

We start with the attractive potential and its integrated form is
\begin{equation}
    \label{eq:lj6}
    U_A = -A\eta_1\eta_2\int\int \frac{d^3\vb{r}_1 d^3\vb{r}_2}{| \vb{r}_1-\vb{r}_2 |^6}~,
\end{equation}
where $\vb{r}_1$ and $\vb{r}_2$ are the position vectors of volume elements, and $\eta_1$ and $\eta_2$ are the number densities of point particles on the two rods. For simplicity, $\eta_1$ and $\eta_2$ are assumed to be constant here.

\subsubsection{Integrated Attractive Potential in 3-Dimensions}

Using the coordinate system in Fig.~\ref{fg:rod-rod-geo}, for thin rods we have $\vb{r}_1 = x\vb{n}_x$ and $\vb{r}_2 = \rho \vb{n}_\rho  + r \vb{n}_z$ and can integrate out the cross sections to get $d^3 \vb{r}_1 = \Sigma_1 dx$ and $d^3 \vb{r}_2 = \Sigma_2 d\rho$, where $\Sigma_1$ and $\Sigma_2$ are the cross-sectional areas of the rods. Furthermore, $| \vb{r}_1-\vb{r}_2 |^2 = r^2+x^2+\rho^2-2x\rho\cos{\theta}$ and the integrated attractive potential becomes
\begin{equation}
    \label{eq:attr_int}
    U_A=-A\int_{\rho_0-l_2}^{\rho_{0}+l_2} \int_{x_{0}-l_1}^{x_{0}+l_1}\frac{ \lambda_{1}\lambda_{2} dxd\rho}{\left( r^2+x^2+\rho ^2-2x\rho \cos\theta \right)^3}~,
\end{equation}
where $l_1$ and $l_2$ are the half lengths, and $\lambda_1 \equiv \eta_1\Sigma_1$ and $\lambda_2 \equiv \eta_2\Sigma_2$ are the line number densities of point particles of the two rods. With a variable change, this integral can be converted into an integral of rational functions, which can be evaluated using the method proposed by Ostrogradski.\cite{Ostrogradski1845a, Ostrogradski1845b} The details are included in the Supporting Information. The final expression of the integrated potential can be written as
\begin{eqnarray}
    \label{eq:int_attr_3D_general}
    U_A &=& -\frac{A\lambda_1\lambda_2}{\sin\theta}
    \left[G\left(x_{0}+l_1,(\rho_{0}+l_2)\sin\theta, r\right)\right. \nonumber \\
    & & -G\left(x_{0}+l_1,(\rho_{0}-l_2)\sin\theta, r\right) \nonumber \\
    & & -G\left(x_{0}-l_1,(\rho_{0}+l_2)\sin\theta, r\right) \nonumber \\
    & & \left. + G\left(x_{0}-l_1,(\rho_{0}-l_2)\sin\theta, r\right)\right]~.
\end{eqnarray}
Here $G(x, a, r)$ is a function defined as
\begin{widetext}
\begin{eqnarray}
    G(x,a,r) &=& \frac{1}{8}\left[\frac{-br^{2} + ax}{r^{2}(a^{2} + r^{2})(b^{2}r^{2} + x^{2})} + \frac{(b^{2}+1)(br^{2}-ax)+2bx^{2}}{(b^{2}r^{2} + x^{2})\left((b^{2} + 1)r^{2} + x^{2}\right)\left(a^{2}(b^{2} + 1) - 2abx + r^{2} + x^{2}\right)} \right. \nonumber \\
    & & + \frac{a(2a^{2} + 3r^{2})}{r^{4}(a^{2} + r^{2})^{\frac{3}{2}}} \arctan\left( \frac{x - ab}{\sqrt{a^{2} + r^{2}}}\right) \nonumber \\
    & & + \frac{x \left( 3 b^{8} r^{8} + b^{6} (3 r^{8} + 11 r^{6} x^{2}) + 3 b^{4} (3 r^{6} x^{2} + 5 r^{4} x^{4}) + 9 b^{2} ( r^{4} x^{4} + r^{2} x^{6}) + 3 r^{2} x^{6} + 2 x^{8} \right) }{r^{4}(b^{2}r^{2} + x^{2})^{3} \left((b^{2} + 1)r^{2} + x^{2}\right)^{\frac{3}{2}} } \times \nonumber \\
    & & \left. \arctan \left( \frac{(b^{2} + 1) a - bx}{\sqrt{(b^{2} + 1)r^{2} + x^{2}}}\right)\right]~,
\end{eqnarray}
\end{widetext}
where $b=\cot\theta$.

For two rods with infinite length, we can adopt $x_0 =\rho_0 =0$ and $l_1=l_2=l$ in Eq.~(\ref{eq:int_attr_3D_general}) and show that
\begin{eqnarray}
    & & \lim_{l\rightarrow \infty} G(l,l\sin\theta,r) \nonumber \\
    &=& \lim_{l\rightarrow \infty} G(-l,-l\sin\theta,r) \nonumber \\
    &=&\frac{1}{2r^4}\arctan \frac{1-\cos\theta}{\sin\theta}~,
\end{eqnarray}
and
\begin{eqnarray}
    & & \lim_{l\rightarrow \infty} G(l,-l\sin\theta,r) \nonumber \\
    &=& \lim_{l\rightarrow \infty} G(-l,l\sin\theta,r) \nonumber \\
    &=&\frac{1}{2r^4}\arctan \frac{1+\cos\theta}{\sin\theta}~.
\end{eqnarray}
Therefore, for two infinitely long rods at separation $|r|$, the integrated mutual attraction is
\begin{equation}
    \label{eq:int_attr_inf_rods}
    \lim_{l\rightarrow \infty} U_A = - \frac{\pi A \lambda_1\lambda_2}{2r^4 \sin\theta}~,
\end{equation}
which agrees with the result obtained by Logan and Tkachenko.\cite{Logan2018}

For crossing rods, $\theta = \pi/2$ and $b=0$. In this case, the expression of  $G(x,a,r)$ can be simplified as
\begin{eqnarray}
    G(x,a,r) &=& \frac{1}{8} \left[
    \frac{ax\left( a^2 + x^2 + 2 r^2\right)}{r^2
    \left( a^2 + r^2\right)
    \left( x^2 + r^2\right)
    \left( a^2 + x^2 + r^2\right)}\right. \nonumber \\
    & & + \frac{a(2a^{2} + 3r^{2})}{r^{4}(a^{2} + r^{2})^{\frac{3}{2}}} \arctan\left( \frac{x}{\sqrt{a^{2} + r^{2}}}\right) \nonumber \\
    & & \left. + \frac{x(2x^{2} + 3r^{2})}{r^{4}(x^{2} + r^{2})^{\frac{3}{2}}} \arctan\left( \frac{a}{\sqrt{x^{2} + r^{2}}}\right)    \right]~,
\end{eqnarray}
which is an odd function of both $x$ and $a$ and is symmetric between $x$ and $a$, as expected. For two crossing rods with the same length, $l_1=l_2=l$, and the common perpendicular passing through their centers, we have $x_0=\rho_0 = 0$ and the relevant function becomes
\begin{eqnarray}
    G(l,l,r) &=& \frac{1}{4} \left[
    \frac{ l^2}{r^2
    \left( l^2 + r^2\right)
    \left( 2 l^2 + r^2\right)}\right. \nonumber \\
    & & \left. + \frac{l(2 l^{2} + 3r^{2})}{r^{4}(l^{2} + r^{2})^{\frac{3}{2}}} \arctan\left( \frac{l}{\sqrt{l^{2} + r^{2}}}\right)   \right]~.
\end{eqnarray}
Using the property $G(l,l,r)= - G(l,-l,r) = -G(-l,l,r) = G(-l,-l,r)$, the integrated attraction becomes
\begin{equation}
    U_A = -4A\lambda_1\lambda_2 G(l,l,r)~.
\end{equation}
In the limit of very long rods, $l\gg |r|$ and we get
\begin{equation}
    \label{eq:int_attr_cro_inf_rods}
    U_A \simeq -A\lambda_1\lambda_2 \left( \frac{1}{2r^2 l^2} + \frac{\pi}{2r^4}\right) \simeq
    - \frac{\pi A\lambda_1 \lambda_2}{2r^4}~,
\end{equation}
which is a known result.\cite{Parsegian2005book} It is also noted that Eq.~(\ref{eq:int_attr_inf_rods}) is reduced to Eq.~(\ref{eq:int_attr_cro_inf_rods}) for crossing rods with infinite length.

\subsubsection{Integrated Attractive Potential in 2-Dimensions}

For two coplanar but nonparallel rods ($r=0$ and $\theta \ne 0$), the integrated attractive potential becomes
\begin{equation}
    U_A =-A\int_{\rho_0-l_2}^{\rho_{0}+l_2} \int_{x_{0}-l_1}^{x_{0}+l_1}\frac{\lambda_{1}\lambda_{2} dxd\rho}{(x^2+\rho ^2-2x\rho \cos\theta)^3}~.
\end{equation}
It can be written as
\begin{eqnarray}
    U_A &=& -\frac{A\lambda_1\lambda_2}{\sin\theta}
    \left[G\left(x_{0}+l_1,(\rho_{0}+l_2)\sin\theta\right) \right. \nonumber \\
    & & -G\left(x_{0}+l_1,(\rho_{0}-l_2)\sin\theta\right) \nonumber \\
    & & -G\left(x_{0}-l_1,(\rho_{0}+l_2)\sin\theta\right) \nonumber \\
    & & \left.
    +G\left(x_{0}-l_1,(\rho_{0}-l_2)\sin\theta\right)\right]~.
\end{eqnarray}
where the function $G(x,a)$ is defined as
\begin{widetext}
\begin{eqnarray}
    G(x,a) &=&\frac{1}{32}\left[ \frac{3}{a^{4}}\arctan\left(\frac{ab-x}{a}\right)  -\frac{3(b^2+1)^2a^{4}-3b(b^2+1)a^{3}x+2a^{2}x^{2}-3bax^{3}+3x^{4}    }{a^{3}x^{3}\left(a^{2}(b^{2}+1)-2bax+x^{2}\right)} \right. \nonumber \\
    & & \left. +\frac{3(b^2+1)^2}{x^{4}}\arctan\left( - \frac{(b^2+1)a-bx}{x}\right) \right]~.
\end{eqnarray}
\end{widetext}

\subsubsection{Integrated Attractive Potential for Parallel Rods}

For two rods parallel with each other ($\theta =0$) and separated at a distance $|r|$, the integrated attractive potential becomes
\begin{equation}
    U_A=-A\int_{\rho_0-l_2}^{\rho_{0}+l_2} \int_{x_{0}-l_1}^{x_{0}+l_1}\frac{\lambda_1\lambda_2 dx d\rho}{\left(r^2+x^2+\rho^2 -2x\rho\right)^3}~.
\end{equation}
The result can be written as
\begin{eqnarray}
    U_A &=& -A\lambda_1\lambda_2
    \left[ K(x_{0}+l_1,\rho_{0}+l_2, r) \right. \nonumber \\
    & & -K(x_{0}+l_1,\rho_{0}-l_2, r) \nonumber \\
    & & -K(x_{0}-l_1,\rho_{0}+l_2, r) \nonumber \\
    & & \left. + K(x_{0}-l_1,\rho_{0}-l_2, r) \right]~.
\end{eqnarray}
where the function $K(x,\rho, r)$ is given by
\begin{eqnarray}
    K(x,\rho,r)& = & \dfrac{1}{8r^2\left[\left(\rho-x\right)^2+r^2\right]} \nonumber \\
    & & -\dfrac{3\left(\rho-x\right)}{8r^5}\arctan\left(\frac{\rho-x}{r}\right)~.
\end{eqnarray}
As expected, the integrated attraction between two parallel rods is a function of $|x_0 - \rho_0|$ (i.e., the offset between the rod centers along their axial direction) and $r$ (i.e., the distance between the axes of the rods), with rod lengths $l_1$ and $l_2$ as parameters. When the offset is 0, the integrated attraction can be simplified to
\begin{eqnarray}
    U_A &=& -\frac{A\lambda_1 \lambda_2}{4r^2} \left[
    \frac{1}{\left( l_2 - l_1 \right)^2+r^2 } - \frac{1}{\left( l_2 + l_1 \right)^2+r^2 } \right. \nonumber \\
    & & 
    -3 \frac{\left( l_2 - l_1 \right)}{r^3} \arctan\left( \frac{l_2 - l_1}{r}\right) \nonumber \\
    & & 
     \left. + 3 \frac{\left( l_2 + l_1 \right)}{r^3} \arctan\left(\frac{l_2 + l_1}{r}\right) \right]~.
\end{eqnarray}
For two rods with the same length ($l_1=l_2=l$), this potential can be further simplified to
\begin{equation}
    U_A = -\frac{A\lambda_1 \lambda_2}{4r^2}
    \left[ \frac{1}{r^2} -\frac{1}{4l^2+r^2}
    + 6\frac{l}{r^3} \arctan \left( \frac{2 l}{r}\right) \right]~.
\end{equation}
For infinitely long rods in a parallel configuration, the attraction per unit length is
\begin{equation}
    \frac{U_A}{2l} = -\frac{3\pi A \lambda_1 \lambda_2}{8 r^5}~,
\end{equation}
which is consistent with the result known for this special geometry.\cite{Parsegian2005book}

\subsubsection{Integrated Attractive Potential for Collinear Rods}

Two rods that are collinear can be represented using the coordinate system in Fig.~\ref{fg:rod-rod-geo} with $r=0$ and $\theta = 0$. That is, the two rods are both on the $x$-axis with their centers located at $x_0$ and $\rho_0$. Of course, since they cannot overlap, we must have $|x_0 - \rho_0|>l_1+l_2$. The integral in Eq.~(\ref{eq:lj6}) can be evaluated and the result is 
\begin{eqnarray}
    U_A &=& -\frac{A\lambda_1\lambda_2}{20}\left[ \left(x_0+l_1 - \rho_0 + l_2\right)^{-4} \right. \nonumber \\
    & & - \left(x_0+l_1 - \rho_0 - l_2\right)^{-4} \nonumber \\
    & & - \left(x_0-l_1 - \rho_0 + l_2\right)^{-4} \nonumber \\
    & & \left. + \left(x_0-l_1 - \rho_0 - l_2\right)^{-4}\right]~.
\end{eqnarray}
As expected, the potential between two collinear rods depends only on $|x_0-\rho_0|$, the distance between their centers, and their respective lengths. For two rods with the same length ($l_1=l_2 =l$), the integrated attraction is reduced to
\begin{eqnarray}
    U_A & =&  -\frac{A\lambda_1\lambda_2}{20}\left[
    \left( x_0 -\rho_0 + 2 l \right)^{-4}
    - 2 \left( x_0 -\rho_0 \right)^{-4} \right. \nonumber \\
    & &  \left.
    + \left( x_0 -\rho_0 - 2 l \right)^{-4}
    \right]~,
\end{eqnarray}
which agrees with the result previously reported.\cite{Parsegian2005book}

\subsection{Repulsive Potential}

In this section, we work out the integrated repulsive potential. The general integrated form of the $1/R^{12}$ repulsion between two bodies can be written as
\begin{equation}
    U_R = B\eta_1\eta_2 \int\int \frac{d^3\vb{r}_1 d^3\vb{r}_2}{| \vb{r}_1-\vb{r}_2 |^{12}}~.
\end{equation}

\subsubsection{Integrated Repulsive Potential in 3-Dimensions}

Using the coordinate system in Fig.~\ref{fg:rod-rod-geo}, for thin rods the integrated repulsive potential becomes
\begin{equation}
    U_R=B\int_{\rho_0-l_2}^{\rho_{0}+l_2} \int_{x_{0}-l_1}^{x_{0}+l_1}\frac{\lambda _{1}\lambda _{2} dxd\rho}{\left( r^2+x^2+\rho ^2-2x\rho \cos\theta \right)^6}~.
\end{equation}
This integral is more challenging than the one in Eq.~(\ref{eq:attr_int}) but can be evaluated similarly using the Ostrogradski method.\cite{Ostrogradski1845a, Ostrogradski1845b} The details can be found in the Supporting Information.

For a general skew 3-dimensional configuration, the repulsive potential after integration can be written as
\begin{eqnarray}
    \label{eq:int_r12_3D}
    U_R &=& \frac{B\lambda_{1}\lambda_{2}}{\sin\theta}\left[
    H\left(x_{0}+l_1,(\rho_{0}+l_2)\sin\theta,r\right) \right. \nonumber \\
    & & - H\left(x_{0}+l_1,(\rho_{0}-l_2)\sin\theta,r\right) \nonumber \\
    & & -H\left(x_{0}-l_1,(\rho_{0}+l_2)\sin\theta,r\right) \nonumber \\
    & & \left. + H\left(x_{0}-l_1,(\rho_{0}-l_2)\sin\theta,r\right)\right]~.
\end{eqnarray}
where $H(x,a,r)$ is a function with its expression included in the Supporting Information.

For two rods with infinite length, we can adopt $x_0 =\rho_0 =0$ and $l_1=l_2=l$ in Eq.~(\ref{eq:int_r12_3D}) and show that
\begin{eqnarray}
    & & \lim_{l\rightarrow \infty} H(l,l\sin\theta,r) \nonumber \\
    &=& \lim_{l\rightarrow \infty} H(-l,-l\sin\theta,r) \nonumber \\
    &=&\frac{1}{5r^{10}}\arctan \frac{1-\cos\theta}{\sin\theta}~,
\end{eqnarray}
and
\begin{eqnarray}
    & & \lim_{l\rightarrow \infty} H(l,-l\sin\theta,r) \nonumber \\
    &=& \lim_{l\rightarrow \infty} H(-l,l\sin\theta,r) \nonumber \\
    &=&\frac{1}{5r^{10}}\arctan \frac{1+\cos\theta}{\sin\theta}~.
\end{eqnarray}
Therefore, for two infinitely long rods at separation $|r|$, the mutual repulsion is
\begin{equation}
    \label{eq:int_rep_inf_rods}
    \lim_{l\rightarrow \infty} U_B = \frac{\pi B \lambda_1\lambda_2}{5r^{10} \sin\theta}~.
\end{equation}

For two crossing rods, $\theta = \pi/2$ and $b=\cot \theta = 0$. The expression of $H(x,a,r)$ in this case can be simplified as
\begin{widetext}
\begin{eqnarray}
    H(x,a,r) &=& \frac{ax}{3840 r^8 (a^2+r^2)^4 (x^2+r^2)^4 (a^2 + x^2 + r^2)^4} \times \left[ 1950 r^{20} \right. \nonumber \\
    & & + r^{18} \left(10440 a^2 + 10440 x^2\right) 
    + r^{16} \left(24660 a^4 + 50260 a^2 x^2 + 24660 x^4 \right) \nonumber \\
    & &  + r^{14} \left(33522 a^6 + 105794 a^4 x^2 + 105794 a^2 x^4 + 33522 x^6 \right) \nonumber \\
    & &  + r^{12} \left(28317 a^8 + 126616 a^6 x^2 + 197508 a^4 x^4 + 126616 a^2 x^6 + 28317 x^8 \right) \nonumber \\
    & & + r^{10} \left(14769 a^{10} + 92579 a^8 x^2 + 207688 a^6 x^4 + 207688 a^4 x^6 + 92579 a^2 x^8 + 14769 x^{10} \right) \nonumber \\ 
    & & + r^8 \left(4365 a^{12} + 40810 a^{10} x^2 + 131185 a^8 x^4 + 189720 a^6 x^6 + 131185 a^4 x^8 + 40810 a^2 x^{10} + 4365 x^{12} \right) \nonumber \\
    & & + r^6 \left(561 a^{14} + 9857 a^{12} x^2 + 48663 a^{10} x^4 + 101399 a^8 x^6 + 101399 a^6 x^8 + 48663 a^4 x^{10} + 9857 a^2 x^{12} + 561 a x^{15}\right) \nonumber \\
    & & + r^4 \left(984 a^{14} x^2 + 9512 a^{12} x^4 + 30344 a^{10} x^6 + 43632 a^8 x^8 + 30344 a^6 x^{10} + 9512 a^4 x^{12} + 984 a^2 x^{14} \right) \nonumber \\
    & & + r^2 \left(720 a^{14} x^4 + 4368 a^{12} x^6 + 9504 a^{10} x^8 + 9504 a^8 x^{10} + 4368 a^6 x^{12} + 720 a^4 x^{14}\right) \nonumber \\
    & & \left. + 192 a^{14} x^6 + 768 a^{12} x^8 + 1152 a^{10} x^{10} + 768 a^8 x^{12} + 192 a^6 x^{14} \right] \nonumber \\
    & & + \frac{315 r^8 a + 840 r^6 a^3 + 1008 r^4 a^5 + 576 r^2 a^7 +128 a^9}{1280 r^{10} \left(a^2+r^2\right)^{9/2}} \times  \arctan\left(\frac{x}{\sqrt{a^2+r^2}} \right) \nonumber\\
    & & + \frac{315 r^8 x + 840 r^6 x^3 + 1008 r^4 x^5 + 576 r^2 x^7 +128 x^9}{1280 r^{10} \left(x^2+r^2\right)^{9/2}} \times  \arctan\left(\frac{a}{\sqrt{x^2+r^2}} \right)~. 
\end{eqnarray}
\end{widetext}

Obviously, in this case $H(x,a,r)$ is an odd function of both $x$ and $a$ and is symmetric between $x$ and $a$, as expected. When the common perpendicular goes through the centers of the two crossing rods with the same length, we have $\theta = \pi/2$, $x_0= \rho_0 = 0$, and $l_1 = l_2 = l$. The integrated repulsion for this system can be expressed in terms of the following function,
\begin{widetext}
\begin{eqnarray}
    \label{eq:rep_cross_rods_sl}
    H(l,l,r) &=& \frac{l^2}{1920 r^8 (l^2+r^2)^4 (2 l^2 + r^2)^4}\times \left(975 r^{12} + 6540 r^{10} l^2 + 17780 r^8 l^4 + 25056 r^6 l^6 + 19648 r^4 l^8 \right. \nonumber \\
    & & \left. + 8448 r^2 l^{10} + 1536 l^{12} \right) + \frac{315 r^8 l + 840 r^6 l^3 + 1008 r^4 l^5 + 576 r^2 l^7 +128 l^9}{640 r^{10} \left(l^2+r^2\right)^{9/2}} \times  \arctan\left(\frac{l}{\sqrt{l^2+r^2}} \right)~. 
\end{eqnarray}
\end{widetext}
Using the property $H(l,l,r)=-H(l,-l,r)=-H(-l,l,r)=H(-l,-l,r)$, the integrated repulsion for two crossing rods of length $2l$ becomes
\begin{equation}
    U_R = 4B\lambda_1\lambda_2 H(l,l,r)~,
\end{equation}
with $H(l,l,r)$ given in Eq.~(\ref{eq:rep_cross_rods_sl}).
For very long rods, $l \gg r$ and the integrated repulsion is
\begin{equation}
    U_R = \frac{\pi B \lambda_1 \lambda_2}{5r^{10}}~.
\end{equation}
As expected, this expression is a special case of Eq.~(\ref{eq:int_rep_inf_rods}) at $\theta = \pi/2$.

\subsubsection{Integrated Repulsive Potential in 2-Dimensions}

The configuration of two unparallel coplanar rods corresponds to $r=0$ and $\theta \ne 0$ in Fig.~\ref{fg:rod-rod-geo}. The integrated repulsive potential can be written as
\begin{eqnarray}
    U_R &=& \frac{B\lambda_{1}\lambda_{2}}{\sin\theta}\left[ H\left(x_{0}+l_1,(\rho_{0}+l_2)\sin\theta\right) \right. \nonumber \\
    & & -H\left(x_{0}+l_1,(\rho_{0}-l_2)\sin\theta\right) \nonumber \\
    & & -H\left(x_{0}-l_1,(\rho_{0}+l_2)\sin\theta\right) \nonumber \\
    & & \left. + H\left(x_{0}-l_1,(\rho_{0}-l_2)\sin\theta\right) \right]~.
\end{eqnarray}
The function $H(x,a)$ has the following form.
\begin{widetext}
\begin{eqnarray}
    H(x,a) &=& -\frac{1}{12800} \left\{\frac{1}{a^9x^9 \left( a^2 \left(b^2+1\right)-2 a b x+x^2 \right)^4} \left[ 315 \left(b^2+1\right)^8 a^{16} \right.\right. \nonumber \\
    & & -2205 b \left(b^2+1\right)^7 a^{15} x + 105 \left(b^2+1\right)^6 \left(63 b^2+11\right) a^{14} x^2 \nonumber \\
    & & - 525 b \left(b^2+1\right)^5 \left(21 b^2+11\right) a^{13} x^3 +21 \left(b^2+1\right)^4 \left(525 b^4+550 b^2+73\right) a^{12} x^4 \nonumber \\
    & & - 21 b \left(b^2+1\right)^3 \left(315 b^4+550 b^2+219\right) a^{11} x^5 \nonumber \\
    & & +3 \left(b^2+1\right)^2 \left(735 b^6+1925 b^4+1533 b^2+279\right) a^{10} x^6\nonumber \\
    & & -3 b \left(b^2+1\right) \left(105 b^6+ 385 b^4+ 511 b^2+279\right) a^9 x^7 +128 a^8 x^8 \nonumber \\
    & & -3 b \left(105 b^6+385 b^4+511 b^2+279\right) a^7 x^9\nonumber \\
    & & + 3 \left(735 b^6+1925 b^4+1533 b^2+279\right) a^6 x^{10} -21 b \left(315 b^4+550 b^2+219\right) a^5 x^{11} \nonumber \\
    & & +21 \left(525 b^4+550 b^2+73\right) a^4 x^{12}-525 b \left(21 b^2+11\right) a^3 x^{13} \nonumber \\
    & & \left. +105 \left(63 b^2+11\right) a^2 x^{14} -2205 b a x^{15}+315 x^{16} \right] \nonumber \\
    & & \left. +\frac{315 \left(b^2+1\right)^5}{x^{10}} \arctan\left(\frac{a (b^2+1)-b x}{x}\right)-\frac{315}{a^{10}} \arctan\left(\frac{ab-x}{a}\right) \right\}~.
\end{eqnarray}
\end{widetext}

 \subsubsection{Integrated Repulsive Potential for Parallel Rods}

For two parallel rods ($\theta = 0$ but $r\ne 0$ in Fig.~\ref{fg:rod-rod-geo}), the integrated repulsive potential adopts the form of
\begin{equation}
    U_R=B\int_{\rho_0-l_2}^{\rho_{0}+l_2} \int_{x_{0}-l_1}^{x_{0}+l_1}\frac{\lambda_1\lambda_2 dx d\rho}{\left(r^2+x^2+\rho^2 -2x\rho\right)^6}
\end{equation}
After integration, the potential can be written as
\begin{eqnarray}
    U_R &=& B\lambda_{1}\lambda_{2}\left[J\left(x_{0}+l_1,\rho_{0}+l_2,r\right) \right. \nonumber \\
    & & - J\left(x_{0}+l_1,\rho_{0}-l_2,r\right) \nonumber \\
    & & -J\left(x_{0}-l_1,\rho_{0}+l_2,r\right) \nonumber \\
    & & \left. +J\left(x_{0}-l_1,\rho_{0}-l_2,r\right) \right]~.
\end{eqnarray}
The function $J(x,\rho,r)$ is given by
\begin{widetext}  
\begin{eqnarray}
    J(x,\rho,r)&=& -\dfrac{63\left( \rho-x\right)}{256r^{11}}\arctan\left(\frac{ \rho-x}{r}\right)+\dfrac{21}{256r^8\left[\left( \rho-x\right)^2+r^2\right]} +\dfrac{21}{640r^6\left[\left( \rho-x\right)^2+r^2\right]^2} \nonumber \\
    & & +\dfrac{3}{160r^4\left[\left( \rho-x\right)^2+r^2\right]^3} + \dfrac{1}{80r^2\left[\left( \rho-x\right)^2+r^2\right]^4}~.
\end{eqnarray}
\end{widetext}
For parallel rods, the integrated repulsion is a function of $|x_0 - \rho_0|$ (the offset between the two rods) only, as expected. When $x_0 = \rho_0$, the offset is 0 and the integrated repulsion between parallel rods becomes
\begin{eqnarray}
    U_R &=& \frac{B\lambda_1 \lambda_2}{8r^2} \left[
    -\dfrac{63\left( l_2-l_1 \right)}{16r^{9}}\arctan\left(\frac{ l_2 - l_1}{r}\right)
    \right. \nonumber \\
   & & + \dfrac{21}{16r^6\left[\left( l_2 - l_1\right)^2+r^2\right]}
    + \dfrac{21}{40r^4\left[\left( l_2 - l_1\right)^2+r^2\right]^2} \nonumber \\
   & & +\dfrac{3}{10r^2\left[\left( l_2 - l_1\right)^2+r^2\right]^3}
   + \dfrac{1}{5\left[\left( l_2 - l_1\right)^2+r^2\right]^4} \nonumber \\
   & &  +\dfrac{63\left( l_2+l_1 \right)}{16r^{9}}\arctan\left(\frac{ l_2 + l_1}{r}\right)
    \nonumber \\
   & & - \dfrac{21}{16r^6\left[\left( l_2 + l_1\right)^2+r^2\right]}
    - \dfrac{21}{40r^4\left[\left( l_2 + l_1\right)^2+r^2\right]^2} \nonumber \\
   & & - \dfrac{3}{10r^2\left[\left( l_2 + l_1\right)^2+r^2\right]^3} \nonumber \\
   & & \left.  - \dfrac{1}{5\left[\left( l_2 + l_1\right)^2+r^2\right]^4}
    \right]~.
\end{eqnarray}
For parallel rods with the same length ($l_1=l_2=l$) and zero offset, the integrated potential can be further reduced to
\begin{eqnarray}
    U_R &=& \frac{B\lambda_1 \lambda_2}{8r^2} \left[ \dfrac{187}{80r^{8}}
+\dfrac{63 l}{8r^{9}}\arctan\left(\frac{ 2l}{r}\right)
   \right. \nonumber \\
   & & - \dfrac{21}{16r^6\left( 4 l^2+r^2\right)}
    - \dfrac{21}{40r^4 \left( 4 l^2+r^2\right)^2} \nonumber \\
   & & \left. - \dfrac{3}{10r^2\left( 4 l^2+r^2\right)^3} 
   - \dfrac{1}{5\left( 4 l^2+r^2\right)^4}
    \right]~.
\end{eqnarray}
For infinitely long rods, the integration repulsion per unit length is
\begin{equation}
    \frac{U_R}{2l} = \frac{63 \pi B\lambda_1 \lambda_2}{256 |r|^{11}}~.
\end{equation}

\subsubsection{Integrated Repulsive Potential for Collinear Rods}

The integrated $1/R^{12}$ repulsion for two collinear rods reads
\begin{eqnarray}
    U_R &=& \frac{B\lambda_1\lambda_2}{110}\left[ \left(x_0+l_1 - \rho_0 + l_2\right)^{-10} \right. \nonumber \\
    & & - \left(x_0+l_1 - \rho_0 - l_2\right)^{-10} \nonumber\\
    & & - \left(x_0-l_1 - \rho_0 + l_2\right)^{-10} \nonumber \\
    & & \left. + \left(x_0-l_1 - \rho_0 - l_2\right)^{-10}\right]~.
\end{eqnarray}
As expected, it is a function of their center-to-center separation, $|x_0-\rho_0|$, only. In this case, $|x_0 - \rho_0| > l_1+l_2$ is required since the two rods cannot overlap.

\subsection{Forces and Torques}

The forces and torques on each rod can be derived from the integrated potentials presented in the previous sections. The results are summarized below. The derivation process is included in the Supporting Information.

\subsubsection{Forces between Rods}

In this section, the subscripts 1 and 2 refer to rod 1 and 2 in Fig.~\ref{fg:rod-rod-geo}, respectively. The subscripts $x$, $\rho$, and $z$ refer to the components of forces and torques along the corresponding axes in the coordinate system defined in Fig.~\ref{fg:rod-rod-geo}.

For rod 1, which is located on the $x$-axis, the total force is
\begin{eqnarray}
    \vb{F}_1 &=& F_{1x}\vb{n}_x+F_{1\rho}\vb{n}_{\rho}+F_{1z}\vb{n}_z \nonumber \\
    &=& -\frac{1}{\sin^2{\theta}}\left(\frac{\partial U}{\partial x_0}+ \cos{\theta} \frac{\partial U}{\partial \rho_0} \right)\vb{n}_x \nonumber \\
    & & + \frac{1}{\sin^2{\theta}} \left(\frac{\partial U}{\partial \rho_0}+ \cos{\theta} \frac{\partial U}{\partial x_0} \right) \vb{n}_{\rho} \nonumber \\
    & & +\frac{\partial U}{\partial r}\vb{n}_z~,
\end{eqnarray}
where $U(x_0,\rho_0,r,\theta)= U_A(x_0,\rho_0,r,\theta) + U_B(x_0,\rho_0,r,\theta)$ is the rod-rod interacting potential obtained in the previous sections. The force on rod 2 can be obtained from the Newton's third law and is $\vb{F}_2 = -\vb{F}_1$.

\subsubsection{Torques between Rods}

For a general skew configurations of two rods as shown in Fig.~\ref{fg:rod-rod-geo}, the torque on rod 1, with respect to its center $(x_0, 0, 0)$, has the following components.
\begin{eqnarray}
    \boldsymbol\tau_{1z} &=& \left[\frac{\partial U}{\partial \theta}-\frac{x_0}{\sin\theta} \left( \frac{\partial U}{\partial \rho_0} +\cos\theta \frac{\partial U}{\partial x_0} \right) \right] \vb{n}_z ~, \nonumber \\
    \boldsymbol\tau_{1y} &=& \left[ -\frac{r}{\sin^2\theta} \left( \frac{\partial U}{\partial x_0} + \cos\theta \frac{\partial U}{\partial \rho_0} \right) + x_0 \frac{\partial U}{\partial r}\right] \vb{n}_y~,
\end{eqnarray}
where $\vb{n}_y = \vb{n}_z \times \vb{n}_x$. The expressions of the force components and the rod-rod potential $U(x_0,\rho_0,r,\theta)$ are given in the previous sections. The components of the torque on rod 2, with respect to its center $(0,\rho_0,r)$, are
\begin{eqnarray}
    \boldsymbol\tau_{2z} &=& \left[-\frac{\partial U}{\partial \theta} +\frac{\rho_0}{\sin\theta} \left( \frac{\partial U}{\partial x_0 } +\cos\theta \frac{\partial U}{\partial \rho_0} \right) \right] \vb{n}_z~, \nonumber \\
    \boldsymbol\tau_{2t} &=& \left[ \frac{r}{\sin^2\theta} \left( \frac{\partial U}{\partial \rho_0 } +\cos\theta \frac{\partial U}{\partial x_0} \right) -\rho_0 \frac{\partial U}{\partial r}  \right] \vb{n}_t~,
\end{eqnarray}
where $\vb{n}_t = \vb{n}_z\times \vb{n}_\rho$.

\section{Theoretical Model of Bead-Rod Interactions}

\subsection{Integrated Bead-Rod Potential}

\begin{figure}[htb]
    \centering
    \includegraphics[width=0.45\textwidth]{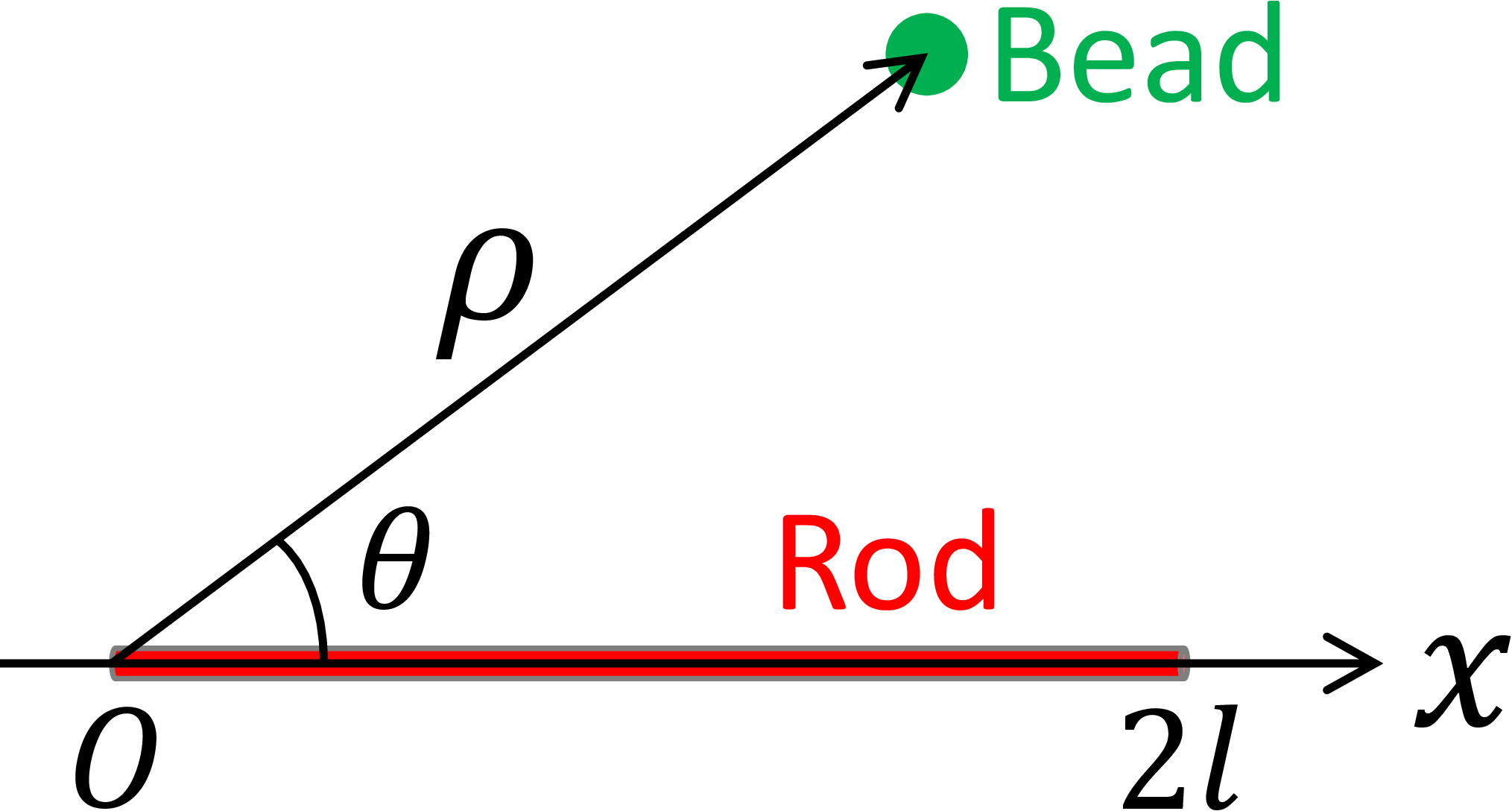}
    \caption{A polar coordinate system describing the configuration of a rod and a point particle (bead).}
    \label{fg:rod-point-geo}
\end{figure}

It is useful to derive the interaction between a point particle and a thin rod by integrating the LJ 12-6 potential. A general bead-rod configuration is sketched in Fig.~\ref{fg:rod-point-geo}. By setting the $x$-axis along the central axis of the rod and choosing one end of the rod as the origin, we can build a polar coordinate system with the point mass located at $(\rho,\theta)$. The interaction potential between the rod and point particle can then be denoted as a function $W(\rho,\theta)$, which represents the integrated form of the LJ 12-6 potential.
\begin{eqnarray}
    W(\rho,\theta) &=& \lambda \int_{0}^{2l} \left[ \frac{B}{\left(x^2 + \rho^2 -2x\rho\cos\theta\right)^6} \right.\nonumber \\
    & & \left. - \frac{A}{\left(x^2 + \rho^2 -2x\rho\cos\theta\right)^3} \right] dx~,
\end{eqnarray}
where $l$ is the half length of the rod and $\lambda$ the line number density of LJ material points that the rod consists of.

In a general case where $\theta \ne 0$ and $\theta \ne \pi$, the integrated bead-rod potential $W(\rho,\theta)$ can be written as
\begin{eqnarray}
    W(\rho,\theta) &=& \lambda\left[L\left(2l,\rho,\theta\right)-L\left(0,\rho,\theta\right) \right]~,
\end{eqnarray}
where the function $L(x,\rho,\theta)$ is
\begin{widetext}
\begin{eqnarray}
    & & L(x,\rho,\theta)\nonumber\\
    &=& \frac{1}{1280 \rho^{10}\sin^2\theta}\left[\frac{128 B \rho^8 (x - \rho \cos\theta)}{(\rho^2 + x^2 - 2 \rho x \cos\theta)^5} + \frac{144 B \rho^6 (x - \rho \cos\theta)}{(\rho^2 + x^2 - 2 \rho x \cos\theta)^4\sin^2\theta} + \frac{168 B \rho^4 (x - \rho \cos\theta) }{(\rho^2 + x^2 - 2 \rho x \cos\theta)^3\sin^4\theta}  \right.
    \nonumber \\
    & & - \frac{10 \rho^2 (x - \rho \cos\theta) \left(32 A \rho^6\sin^6\theta - 21 B \right)}{(\rho^2 + x^2 - 2 \rho x \cos\theta)^2\sin^6\theta} - \frac{15 (x - \rho \cos\theta)\left(32 A \rho^6 \sin^6\theta - 21 B \right)}{(\rho^2 + x^2 - 2 \rho x \cos\theta)\sin^8\theta}\nonumber \\
    & & \left. - \frac{15 \arctan\left(\frac{x - \rho \cos\theta}{\rho|\sin\theta|}\right)\left(32 A \rho^6 \sin^6\theta - 21 B \right)}{\rho|\sin\theta|^9}\right]~.
\end{eqnarray}
\end{widetext}

For the special situation where the point particle and rod are collinear ($\theta = 0$ or $\theta = \pi$), the integrated bead-rod potential can be expressed as $W(\rho_s)$, which has the following form.
\begin{eqnarray}
    W(\rho_s) &=& \lambda\left[-\frac{A}{5 (\rho_s+2l)^5} -\frac{A}{5 \rho_s^5} \right.\nonumber \\
    & & \left. +\frac{B}{11 (\rho_s+2l)^{11}} +\frac{B}{11 \rho_s^{11}}\right]~.
\end{eqnarray}
Here $\rho_s$ is the distance of the point particle from the near end of the rod, which is obviously the shortest distance between the point particle and rod. Using the coordinate system sketched in Fig.~\ref{fg:rod-point-geo}, $\rho_s = \rho-2l$ when $\theta =0$ while $\rho_s = \rho$ when $\theta =\pi$.

\subsection{Forces and Torques in Bead-Rod Interactions}

Using the coordinate system defined in Fig.~\ref{fg:rod-point-geo}, the force on the bead (i.e., the point particle) is
\begin{equation}
    \vb{F}_{2} = \frac{1}{\rho \sin\theta}\frac{\partial W}{ \partial \theta} \vb{n}_x - \left( \frac{\partial W}{\partial \rho} + \frac{\cos \theta}{\rho \sin\theta}\frac{\partial W}{ \partial \theta} \right) \vb{n}_\rho~.
\end{equation}
The force on the rod can be obtained from the Newton's third law as $\vb{F}_1 = -\vb{F}_2$.

The torque on the rod is
\begin{equation}
    \boldsymbol{\tau}_{1z} = \left[ -l \sin\theta \frac{\partial W}{\partial \rho} + \left( 1- \frac{l}{\rho} \cos\theta \right) \frac{\partial W}{\partial \theta} \right]\vb{n}_z~,
\end{equation}
where $\vb{n}_z \equiv \frac{1}{\sin\theta} \vb{n}_x\times \vb{n}_\rho$ is the unit vector perpendicular to the $x\rho$ plane. As expected, the torque is 0 for $\theta =0$ and $\pi$ where the point particle and rod are collinear. This is evident from the fact that at a given $\rho$, the potential $W(\rho,\theta)$ achieves extremes at $\theta =0$ and $\pi$. It can also be proven that the torque on the rod is 0 when $\rho = l/\cos\theta$, i.e., when the point particle is on the perpendicular bisector of the rod.

\section{Conclusions}

Using the Ostrogradski method, we have successfully integrated the Lennard-Jones (LJ) 12-6 potential for two thin rods of both finite and infinite lengths in arbitrary 3-dimensional configurations, with each rod modeled as a continuous distribution of LJ point particles. Both integrated attraction and repulsion are expressed in analytical forms. The previously known expressions for two thin rods in special configurations (e.g., parallel or crossing) are all asymptotic cases of the general forms reported here. The general closed forms can greatly facilitate theoretical treatments and computational studies of systems involving rod-like objects. Examples include rod-shaped liquid crystal molecules, colloidal nanorods, polymer segments, nanowires, carbon nanotubes, rod-like microbes, filamentous materials, and filament bundles. An analytical form is also obtained for the integrated LJ potential between a point particle and a thin rod. Such potential can be used to investigate the behavior of thin rods in a solvent modeled as a liquid consisting of LJ beads.

\section*{Acknowledgements}
This material is based upon work supported by the Air Force Office of Scientific Research under award number FA9550-18-1-0433. Any opinions, findings and conclusions or recommendations expressed in this material are those of the author(s) and do not necessarily reflect the views of the U.S. Department of Defense.


%

\clearpage
\newpage
\onecolumngrid
\renewcommand{\thesection}{S\arabic{section}}
\setcounter{section}{0}
\renewcommand{\thefigure}{S\arabic{figure}}
\setcounter{figure}{0}
\renewcommand{\theequation}{S\arabic{equation}}
\setcounter{equation}{0}
\renewcommand{\thepage}{SI-\arabic{page}}
\setcounter{page}{1}
\begin{center}
{\bf \large Supporting Information for ``Analytical Interaction Potential for Lennard-Jones Rods''}
\end{center}
\begin{center}
{\bf Junwen Wang$^{2,4,5}$, Gary Seidel$^{3,2,5}$, and Shengfeng Cheng$^{1,2,4,5}$}
\end{center}
\begin{center}
{$^1$Department of Physics, $^2$Department of Mechanical Engineering, $^3$Department of Aerospace and Ocean Engineering, $^4$Center for Soft Matter and Biological Physics, $^5$Macromolecules Innovation Institute,\\ Virginia Tech, Blacksburg 24061, USA}
\end{center}


In this Supporting Information, we first show how the function $H(x,a,r)$, which is involved in the expression of the integrated repulsive potential between thin rods, is derived. Then we show how the expressions of forces and torques can be derived from the integrated potentials. We also include comparisons between the analytical results and the numerical results from direct numerical evaluation of the double integrals for the integrated potentials. The comparison fully verifies the analytical formulae presented in the main text.

\section{Integrated Repulsive Potential}

Using the coordinate system in Fig.~1 of the main text, the integrated repulsive potential between two thin rods has the following form.
\begin{eqnarray}
\label{eq:rep_integral}
    U_R &=& B\lambda_{1}\lambda_{2}\int_{\rho_0-l_2}^{\rho_{0}+l_2} \int_{x_{0}-l_1}^{x_{0}+l_1}\frac{dxd\rho}{\left(r^2+x^2+\rho ^2-2x\rho \cos\theta\right)^6} \nonumber \\
    &=& B\lambda_{1}\lambda_{2}\int_{\rho_0-l_2}^{\rho_{0}+l_2} \int_{x_{0}-l_1}^{x_{0}+l_1}\frac{dxd\rho}{\left[r^2+(x-\rho \cos\theta)^2+\rho^2 \sin^2\theta \right]^6}
\end{eqnarray}
We introduce the changes of variables: $a=\rho \sin\theta$ and $y=x-\rho \cos\theta$. Then $da=d\rho \sin\theta$ and $d\rho =\frac{da}{\sin\theta}$. Note that $\rho \cos\theta =\rho \sin\theta \frac{\cos\theta}{\sin\theta}=a \cot\theta$. If we set $b=\cot \theta$, then $\rho\cos\theta = ab$ and $y=x-ab$. At a constant $\rho$, we have $dy = dx$ and the integration over $x$ in Eq.~(\ref{eq:rep_integral}) can be completed and the resulting integrated repulsive potential becomes
\begin{eqnarray}
    U_R &=& \frac{B\lambda_{1}\lambda_{2}}{\sin\theta}\int_{(\rho_0-l_2)\sin\theta}^{(\rho_{0}+l_2)\sin\theta} da \left[ \frac{63}{256(r^2+a^2)^{11/2}}\arctan(\frac{y}{\sqrt{r^2+a^2}})\right.\nonumber \\
    & & +\frac{193y}{256(r^2+a^2)(r^2+a^2+y^2)^5}+\frac{63y^9}{256(r^2+a^2)^{5}(r^2+a^2+y^2)^5}\nonumber \\
    & & +\frac{147y^7}{128(r^2+a^2)^4(r^2+a^2+y^2)^5}+\frac{21y^5}{10(r^2+a^2)^3(r^2+a^2+y^2)^5}\nonumber \\
    & & \left. +\frac{237y^3}{128(r^2+a^2)^2(r^2+a^2+y^2)^5}\right]^{y=x_{0}+l_1-ab}_{y=x_{0}-l_1-ab}
\end{eqnarray}

If we define
\begin{eqnarray}
    H(x,a,r)&=&\int da \left[ \frac{63}{256(r^2+a^2)^{11/2}}\arctan(\frac{y}{\sqrt{r^2+a^2}})\right.\nonumber \\
    & & +\frac{193y}{256(r^2+a^2)(r^2+a^2+y^2)^5}+\frac{63y^9}{256(r^2+a^2)^{5}(r^2+a^2+y^2)^5}\nonumber \\
    & & +\frac{147y^7}{128(r^2+a^2)^4(r^2+a^2+y^2)^5}+\frac{21y^5}{10(r^2+a^2)^3(r^2+a^2+y^2)^5}\nonumber \\
    & & \left. +\frac{237y^3}{128(r^2+a^2)^2(r^2+a^2+y^2)^5}\right]
\end{eqnarray}
where $y=x-ab$, then $U_R$ can be written as
\begin{eqnarray}
    U_R &=& \frac{B\lambda_{1}\lambda_{2}}{\sin\theta} \left[H(x_{0}+l_1,(\rho_{0}+l_2)\sin\theta,r)-H(x_{0}+l_1,(\rho_{0}-l_2)\sin\theta,r)\right.\nonumber \\
    & & \left. -H(x_{0}-l_1,(\rho_{0}+l_2)\sin\theta,r)+H(x_{0}-l_1,(\rho_{0}-l_2)\sin\theta,r)\right]
\end{eqnarray}
The target integral to compute is
\begin{eqnarray}
    H(x,a,r) &=& \int da \left[ \frac{63}{256(r^2+a^2)^{11/2}}\arctan(\frac{x-ab}{\sqrt{r^2+a^2}})\right.\nonumber\\
    & & +\frac{193(x-ab)}{256(r^2+a^2)\left[r^2+a^2+(x-ab)^2\right]^5}+\frac{63(x-ab)^9}{256(r^2+a^2)^{5}\left[r^2+a^2+(x-ab)^2\right]^5}\nonumber\\
    & & +\frac{147(x-ab)^7}{128(r^2+a^2)^4\left[r^2+a^2+(x-ab)^2\right]^5}+\frac{21(x-ab)^5}{10(r^2+a^2)^3\left[r^2+a^2+(x-ab)^2\right]^5}\nonumber\\
    & & \left.+\frac{237(x-ab)^3}{128(r^2+a^2)^2\left[r^2+a^2+(x-ab)^2\right]^5}\right]
\end{eqnarray}
The computation of $H(x,a,r)$ can be broken down to the calculation of the following six integrals, labeled as \#1 to \#6.
\begin{eqnarray}
\label{eq:Hfunc}
    H(x,a,r)&=& \int \frac{193(x-ab)}{256(r^2+a^2)\left[r^2+a^2+(x-ab)^2\right]^5}da~~~~~~~~~~(\#1) \nonumber\\
    & & + \int \frac{63(x-ab)^9}{256(r^2+a^2)^{5}\left[r^2+a^2+(x-ab)^2\right]^5}da ~~~~~~~~~~(\#2) \nonumber\\
    & & + \int \frac{147(x-ab)^7}{128(r^2+a^2)^4\left[r^2+a^2+(x-ab)^2\right]^5}da ~~~~~~~~~~(\#3) \nonumber \\
    & & + \int \frac{21(x-ab)^5}{10(r^2+a^2)^3\left[r^2+a^2+(x-ab)^2\right]^5}da ~~~~~~~~~~(\#4) \nonumber \\
    & & + \int \frac{237(x-ab)^3}{128(r^2+a^2)^2\left[r^2+a^2+(x-ab)^2\right]^5}da ~~~~~~~~~~(\#5) \nonumber \\
    & & + \int \frac{63}{256(r^2+a^2)^{11/2}}\arctan(\frac{x-ab}{\sqrt{r^2+a^2}}) da ~~~~~~~~~~(\#6) 
\end{eqnarray}

\subsection{Integral \#1}

As shown in Eq.~(\ref{eq:Hfunc}), the function $H(x,a,r)$ is the sum of six integrals, five of which are in the form of an integral of a rational function. These integrals can be computed with Ostrogradski's integration method \cite{Ostrogradski1845a,Ostrogradski1845b}. Here we demonstrate the process using the first integral in Eq.~(\ref{eq:Hfunc}), the integral \#1, as an example.
The integral to compute is
\begin{equation}
    \int \frac{x-ab}{(r^2+a^2)\left[r^2+a^2+(x-ab)^2\right]^5}da
\end{equation}
We define
\begin{equation}
    P(a)=x-ab
\end{equation}
and
\begin{equation}
    Q(a)=(r^2+a^2)\left[r^2+a^2+(x-ab)^2\right]^5
\end{equation}
then
\begin{equation}
    Q^\prime(a)=10 \left[(b^2+1)a-bx\right] ( r^2+a^2) \left[r^2 + a^2 + (x-ab)^2\right]^4 + 2 a \left[r^2+a^2 + (x-ab)^2\right]^5
\end{equation}
The greatest common divisor of $Q(a)$ and $Q^\prime(a)$ is
\begin{equation}
    Q_{1}(a)=\left[r^2 + a^2 + (x-ab)^2\right]^4
\end{equation}
and dividing $Q(a)$ by $Q_{1}(a)$ yields
\begin{equation}
    Q_{2}(a)=(r^2+a^2)\left[r^2+a^2+(x-ab)^2\right]
\end{equation}
Using Ostrogradski's method \cite{Ostrogradski1845a,Ostrogradski1845b},
\begin{eqnarray}
\label{eq:Ostro_transform}
    & & \int \frac{x-ab}{(r^2+a^2)\left[r^2+a^2+(x-ab)^2\right]^5}da \nonumber \\
    &=& \frac{\mathcal{A}a^7+\mathcal{B}a^6+\mathcal{C}a^5+\mathcal{D}a^4+\mathcal{E}a^3+\mathcal{F}a^2+\mathcal{G}a+\mathcal{H}}{Q_{1}(a)} + \int \frac{\mathcal{R}a^3+\mathcal{S}a^2+\mathcal{T}a+\mathcal{V}}{Q_{2}(a)}da \nonumber\\
    &=& \frac{\mathcal{A}a^7+\mathcal{B}a^6+\mathcal{C}a^5+\mathcal{D}a^4+\mathcal{E}a^3+\mathcal{F}a^2+\mathcal{G}a+\mathcal{H}}{\left[r^2 +a^2+ (x-ab)^2\right]^4} + \int \frac{\mathcal{R}a^3+\mathcal{S}a^2+\mathcal{T}a+\mathcal{V}}{(r^2+a^2)\left[r^2+a^2+(x-ab)^2\right]}da
\end{eqnarray}
Here all capital letters in the cursive font are functions of $x$, $b$,and $r$. That is
\begin{equation}
    \mathcal{A}=\mathcal{A}(x,b,r),~\mathcal{B}=\mathcal{B}(x,b,r),~...,~\mathcal{V}=\mathcal{V}(x,b,r)
\end{equation}
Taking derivatives with respect to $a$ on both sides of Eq.~(\ref{eq:Ostro_transform}) and equating the corresponding coefficients yields the following linear equation,
\begin{eqnarray}
\label{eq:linear_system}
    x-ab &=& (7\mathcal{A}a^6+6\mathcal{B}a^5+5\mathcal{C}a^4+4\mathcal{D}a^3+3\mathcal{E}a^2+2\mathcal{F}a+\mathcal{G})(r^2+a^2)\left[r^2+a^2+(x-ab)^2\right]\nonumber\\
    & & -4\left[2a-2b(x-ab)\right](\mathcal{A}a^7+\mathcal{B}a^6+\mathcal{C}a^5+\mathcal{D}a^4+\mathcal{E}a^3+\mathcal{F}a^2+\mathcal{G}a+\mathcal{H})(r^2+a^2)\nonumber\\
    & & +(\mathcal{R}a^3+\mathcal{S}a^2+\mathcal{T}a+\mathcal{V})\left[r^2+a^2+(x-ab)^2\right]^4
\end{eqnarray}
The Python code used to solve this equation symbolically is attached as a separate file.

The integral term on the right hand side of Eq.~(\ref{eq:Ostro_transform}) can be written as
\begin{eqnarray}
    & & \int \frac{\mathcal{R}a^3+\mathcal{S}a^2+\mathcal{T}a+\mathcal{V}}{(r^2+a^2)\left[r^2+a^2+(x-ab)^2\right]}da \nonumber\\
    &=& \int \frac{\mathcal{R}a^3}{(r^2+a^2)\left[r^2+a^2+(x-ab)^2\right]}da + \int \frac{\mathcal{S}a^2}{(r^2+a^2)\left[r^2+a^2+(x-ab)^2\right]}da\nonumber \\
    & & \int \frac{\mathcal{T}a}{(r^2+a^2)\left[r^2+a^2+(x-ab)^2\right]}da + \int \frac{\mathcal{V}}{(r^2+a^2)\left[r^2+a^2+(x-ab)^2\right]}da
\end{eqnarray}
Below we solve each of these four integrals separately. For the first integral, performing a partial fraction decomposition and using linearity, we obtain
\begin{eqnarray}
\label{eq:first_int}
    & & \int \frac{a^3}{(r^2+a^2)\left[r^2+a^2+(x-ab)^2\right]}da\nonumber\\
    &=& \int \left[\frac{\left(x^{4}+\left(3 b^{2}+1\right) r^{2} x^{2}-b^{2} r^{4}\right) a-2 b r^{2} x^{3}-2 b r^{4} x}{\left(x^{4}+2 b^{2} r^{2} x^{2}+b^{4} r^{4}\right)\left(\left(b^{2}+1\right) a^{2}-2 b x a+x^{2}+r^{2}\right)} \right. \nonumber\\
    & & \left. -\frac{\left(r^{2} x^{2}-b^{2} r^{4}\right) a-2 b r^{4} x}{\left(x^{4}+2 b^{2} r^{2} x^{2}+b^{4} r^{4}\right)\left(a^{2}+r^{2}\right)}\right]da \nonumber \\
    &=& \frac{1}{x^{4}+2 b^{2} r^{2} x^{2}+b^{4} r^{4}} \int \frac{\left(x^{4}+\left(3 b^{2}+1\right) r^{2} x^{2}-b^{2} r^{4}\right) a-2 b r^{2} x^{3}-2 b r^{4} x}{\left(b^{2}+1\right) a^{2}-2 b x a+x^{2}+r^{2}} da\nonumber \\
    & & -\frac{r^{2}}{x^{4}+2 b^{2} r^{2} x^{2}+b^{4} r^{4}} \int \frac{\left(x^{2}-b^{2} r^{2}\right) a-2 b r^{2} x}{a^{2}+r^{2}} da
\end{eqnarray}

We now solve the following integral.
\begin{equation}
\label{eq:integral_S16}
    \int \frac{\left(x^{4}+\left(3 b^{2}+1\right) r^{2} x^{2}-b^{2} r^{4}\right) a-2 b r^{2} x^{3}-2 b r^{4} x}{\left(b^{2}+1\right) a^{2}-2 b x a+x^{2}+r^{2}} da
\end{equation}
Using the following identity,
\begin{eqnarray}
    & & \left(x^{4}+\left(3 b^{2}+1\right) r^{2} x^{2}-b^{2} r^{4}\right) a-2 b r^{2} x^{3}-2 b r^{4} x \nonumber\\
    &=& \frac{x^{4}+\left(3 b^{2}+1\right) r^{2} x^{2}-b^{2} r^{4}}{2\left(b^{2}+1\right)}\left(2\left(b^{2}+1\right) a-2 b x\right)\nonumber\\
    & & +\frac{b x\left(x^{4}+\left(3 b^{2}+1\right) r^{2} x^{2}-b^{2} r^{4}\right)}{b^{2}+1}-2 b r^{2} x^{3}-2 b r^{4} x
\end{eqnarray}
the integral in Eq.~(\ref{eq:integral_S16}) can be split as
\begin{eqnarray}
\label{eq:split}
    & & \int \frac{\left(x^{4}+\left(3 b^{2}+1\right) r^{2} x^{2}-b^{2} r^{4}\right) a-2 b r^{2} x^{3}-2 b r^{4} x}{\left(b^{2}+1\right) a^{2}-2 b x a+x^{2}+r^{2}} da\nonumber\\
    &=& \int\left[\frac{\left(x^{4}+\left(3 b^{2}+1\right) r^{2} x^{2}-b^{2} r^{4}\right)\left(2\left(b^{2}+1\right) a-2 b x\right)}{2\left(b^{2}+1\right)\left(\left(b^{2}+1\right) a^{2}-2 b x a+x^{2}+r^{2}\right)}\right.\nonumber\\
    & & \left.+\frac{\frac{b x\left(x^{4}+\left(3 b^{2}+1\right) r^{2} x^{2} b^{2} r^{2}\right)}{b^{2}+1}-2 b r^{2} x^{3}-2 b r^{4} x}{\left(b^{2}+1\right) a^{2}-2 b x a+x^{2}+r^{2}}\right] da \nonumber\\
    &=& \frac{x^{4}+\left(3 b^{2}+1\right) r^{2} x^{2}-b^{2} r^{4}}{b^{2}+1} \int \frac{\left(b^{2}+1\right) a-b x}{\left(b^{2}+1\right) a^{2}-2 b x a+x^{2}+r^{2}} da\nonumber\\
    & & + \left[\frac{b x\left(x^{4}+\left(3 b^{2}+1\right) r^{2} x^{2}-b^{2} r^{4}\right)}{b^{2}+1}-2 b r^{2} x^{3}-2 b r^{4} x\right] \times \nonumber\\
    & & \int \frac{1}{\left(b^{2}+1\right) a^{2}-2 b x a+x^{2}+r^{2}} da
\end{eqnarray}

We now need to solve the following integral
$$\int \frac{\left(b^{2}+1\right) a-b x}{\left(b^{2}+1\right) a^{2}-2 b x a+x^{2}+r^{2}} da$$
Using the substitution
\begin{equation}
    u=(b^2+1)a^2-2bxa+x^2+r^2
\end{equation}
we have
\begin{equation}
    \frac{du}{da}=2(b^2+1)a-2bx
\end{equation}
and
\begin{equation}
    da=\frac{1}{2(b^2+1)a-2bx}du
\end{equation}
Therefore
\begin{equation}
    \int \frac{\left(b^{2}+1\right) a-b x}{\left(b^{2}+1\right) a^{2}-2 b x a+x^{2}+r^{2}} da = \frac{1}{2}\int \frac{1}{u}du=\frac{1}{2} \ln u
\end{equation}
That is
\begin{equation}
    \int \frac{\left(b^{2}+1\right) a-b x}{\left(b^{2}+1\right) a^{2}-2 b x a+x^{2}+r^{2}} da = \frac{1}{2} \ln \left[(b^2+1)a^2-2bxa+x^2+r^2\right]
\end{equation}

The integral in the second term on the right hand side of the final expression in Eq.~(\ref{eq:split}) can be rewritten as
\begin{eqnarray}
    \label{eq:split_second_term}
    & & \int \frac{1}{\left(b^{2}+1\right) a^{2}-2 b x a+x^{2}+r^{2}} da \nonumber\\
    &=& \int \frac{1}{\left(\sqrt{b^{2}+1} a-\frac{b x}{\sqrt{b^{2}+1}}\right)^{2}-\frac{b^{2} x^{2}}{b^{2}+1}+x^{2}+r^{2}} da
\end{eqnarray}
Using the following substitution
\begin{equation}
    w=\frac{\left(b^{2}+1\right) a-b x}{\sqrt{b^{2}+1} \sqrt{-\frac{b^{2} x^{2}}{b^{2}+1}+x^{2}+r^{2}}}
\end{equation}
we get
\begin{equation}
    \frac{dw}{da}=\frac{\sqrt{b^{2}+1}}{\sqrt{-\frac{b^{2} x^{2}}{b^{2}+1}+x^{2}+r^{2}}} 
\end{equation}
and
\begin{equation}
    da=\frac{\sqrt{-\frac{b^{2} x^{2}}{b^{2}+1}+x^{2}+r^{2}}}{\sqrt{b^{2}+1}} dw
\end{equation}
The integral in Eq.~(\ref{eq:split_second_term}) becomes
\begin{eqnarray}
    & & \int \frac{1}{\left(\sqrt{b^{2}+1} a-\frac{b x}{\sqrt{b^{2}+1}}\right)^{2}-\frac{b^{2} x^{2}}{b^{2}+1}+x^{2}+r^{2}} da \nonumber\\
    &=& \int \frac{\sqrt{-\frac{b^{2} x^{2}}{b^{2}+1}+x^{2}+r^{2}}}{\sqrt{b^{2}+1}\left(\left(-\frac{b^{2} x^{2}}{b^{2}+1}+x^{2}+r^{2}\right) w^{2}-\frac{b^{2} x^{2}}{b^{2}+1}+x^{2}+r^{2}\right)} dw \nonumber\\
    &=& \frac{1}{\sqrt{b^{2}+1} \sqrt{-\frac{b^{2} x^{2}}{b^{2}+1}+x^{2}+r^{2}}} \int \frac{1}{w^{2}+1} dw\nonumber\\
    &=& \frac{1}{\sqrt{b^{2}+1} \sqrt{-\frac{b^{2} x^{2}}{b^{2}+1}+x^{2}+r^{2}}}\arctan w \nonumber\\
    &=& \frac{1}{\sqrt{b^{2}+1} \sqrt{-\frac{b^{2} x^{2}}{b^{2}+1}+x^{2}+r^{2}}}\arctan(\frac{\left(b^{2}+1\right) a-b x}{\sqrt{b^{2}+1} \sqrt{-\frac{b^{2} x^{2}}{b^{2}+1}+x^{2}+r^{2}}})\nonumber\\
    &=& \frac{1}{\sqrt{(b^2+1)r^{2}+x^2}}\arctan(\frac{\left(b^{2}+1\right) a-b x}{\sqrt{(b^2+1)r^{2}+x^2}})
\end{eqnarray}

So the integral in Eq.~(\ref{eq:split}) becomes
\begin{eqnarray}
    & & \frac{x^{4}+\left(3 b^{2}+1\right) r^{2} x^{2}-b^{2} r^{4}}{b^{2}+1} \int \frac{\left(b^{2}+1\right) a-b x}{\left(b^{2}+1\right) a^{2}-2 b x a+x^{2}+r^{2}} d a \nonumber\\
    & & +\left[\frac{b x\left(x^{4}+\left(3 b^{2}+1\right) r^{2} x^{2}-b^{2} r^{4}\right)}{b^{2}+1}-2 b r^{2} x^{3}-2 b r^{4} x\right]\times\nonumber\\
    & & \int \frac{1}{\left(b^{2}+1\right) a^{2}-2 b x a+x^{2}+r^{2}} da \nonumber\\
    &=& \frac{x^{4}+\left(3 b^{2}+1\right) r^{2} x^{2}-b^{2} r^{4}}{2\left(b^{2}+1\right)} \ln \left[\left(b^{2}+1\right) a^{2}-2 b x a+x^{2}+r^{2}\right]\nonumber\\
    & & + \frac{\frac{b x\left(x^{4}+\left(3 b^{2}+1\right) r^{2} x^{2} b^{2} r^{2}\right)}{b^{2}+1}-2 b r^{2} x^{3}-2 b r^{4} x}{\sqrt{(b^2+1)r^{2}+x^2}}\arctan(\frac{\left(b^{2}+1\right) a-b x}{\sqrt{(b^2+1)r^{2}+x^2}})
\end{eqnarray}

The integral in the second term on the right hand side of the final expression in Eq.~(\ref{eq:first_int}) can be completed as
\begin{eqnarray}
    \int \frac{\left(x^{2}-b^{2} r^{2}\right) a-2 b r^{2} x}{a^{2}+r^{2}} da &=& (x^2-b^2r^2)\int \frac{a}{a^2+r^2}da-2br^2x\int \frac{1}{a^2+r^2}da \nonumber\\
    &=& \frac{x^2-b^2r^2}{2} \ln(a^2+r^2) - 2brx \arctan(\frac{a}{r})
\end{eqnarray}

Putting all things together, the result of the target integral in Eq.~(\ref{eq:first_int}) is
\begin{eqnarray}
    & & \int \frac{a^3}{(r^2+a^2)\left[r^2+a^2+(x-ab)^2\right]}da\nonumber\\
    &=& \frac{1}{2\left(b^{2} r^{2}+x^{2}\right)^{2}} \left[ \left(b^{2} r^{4}-r^{2} x^{2}\right) \ln \left(a^{2}+r^{2}\right)\right.\nonumber\\
    & & +\frac{x^{2}\left(r^{2}+x^{2}\right)-b^{2}\left(r^{4}-3 r^{2} x^{2}\right)}{b^{2}+1} \ln \left(a^{2}\left(b^{2}+1\right)-2 a b x+r^{2}+x^{2}\right) \nonumber\\
    & & + \frac{2 b x\left(-\left(3 b^{2}+2\right) r^{4}+\left(b^{2}-1\right) r^{2} x^{2}+x^{4}\right)}{\left(b^{2}+1\right) \sqrt{\left(b^{2}+1\right) r^{2}+x^{2}}} \arctan\left(\frac{a b^{2}+a-b x}{\sqrt{\left(b^{2}+1\right) r^{2}+x^{2}}}\right) \nonumber\\
    & & \left. + 4 b r^{3} x \arctan\left(\frac{a}{r}\right)\right]
\end{eqnarray}

Using the same method, we obtain
\begin{eqnarray}
    & & \int \frac{a^2}{(r^2+a^2)\left[r^2+a^2+(x-ab)^2\right]}da\nonumber\\
    &=& \frac{1}{\left(b^{2} r^{2}+x^{2}\right)^{2}} \left[ b r^{2} x \ln \frac{a^{2}\left(b^{2}+1\right)-2 a b x+r^{2}+x^{2}}{a^{2}+r^{2}}\right.\nonumber\\
    & & + r\left(b^{2} r^{2}-x^{2}\right) \arctan\left(\frac{a}{r}\right) \nonumber \\
    & & \left. + \frac{b^{2}\left(r^{2} x^{2}-r^{4}\right)+x^{2}\left(r^{2}+x^{2}\right)}{\sqrt{\left(b^{2}+1\right) r^{2}+x^{2}}} \arctan\left(\frac{a b^{2}+a-b x}{\sqrt{\left(b^{2}+1\right) r^{2}+x^{2}}}\right) \right]
\end{eqnarray}
and
\begin{eqnarray}
    & & \int \frac{a}{(r^2+a^2)\left[r^2+a^2+(x-ab)^2\right]}da\nonumber\\
    &=& \frac{1}{2\left(b^{2} r^{2}+x^{2}\right)^{2}} \left[ \left(b^{2} r^{2}-x^{2}\right) \ln \frac{a^{2}\left(b^{2}+1\right)-2 a b x+r^{2}+x^{2}}{a^{2}+r^{2}}\right.\nonumber\\
    & & -4 b r x \arctan\left(\frac{a}{r}\right)\nonumber\\
    & & \left. + \frac{2 b x\left(\left(b^{2}+2\right) r^{2}+x^{2}\right)}{\sqrt{\left(b^{2}+1\right) r^{2}+x^{2}}}\arctan\left(\frac{a b^{2}+a-b x}{\sqrt{\left(b^{2}+1\right) r^{2}+x^{2}}}\right) \right]
\end{eqnarray}
and
\begin{eqnarray}
    & & \int \frac{1}{(r^2+a^2)\left[r^2+a^2+(x-ab)^2\right]}da\nonumber\\
    &=& \frac{1}{\left(b^{2} r^{2}+x^{2}\right)^{2}} \left[ -b x \ln \frac{a^{2}\left(b^{2}+1\right)-2 a b x+r^{2}+x^{2}}{a^{2}+r^{2}}\right.\nonumber\\
    & & -\frac{b^{2} r^{2}-x^{2}}{r}\arctan\left(\frac{a}{r}\right)\nonumber\\
    & & \left.+\frac{b^{4} r^{2}+b^{2}\left(r^{2}+x^{2}\right)-x^{2}}{\sqrt{\left(b^{2}+1\right) r^{2}+x^{2}}}\arctan\left(\frac{a b^{2}+a-b x}{\sqrt{\left(b^{2}+1\right) r^{2}+x^{2}}}\right)\right]
\end{eqnarray}
The integral in Eq.~(\ref{eq:Ostro_transform}) is then completed after we obtain $R$, $S$, $T$, and $U$ using linear equations such as the one in Eq.~(\ref{eq:linear_system}).

The final result for the integral \#1 is
\begin{eqnarray}
    & & \int \frac{(x-ab)}{(r^2+a^2)\left[ r^2+a^2+(x-ab)^2\right]^5}da\nonumber\\
    &=&\frac{r^2 b^3-a x b^2+\left(r^2+2 x^2\right) b-a x}{8 \left(b^2 r^2+x^2\right) \left[\left(b^2+1\right) r^2+x^2\right] \left[\left(b^2+1\right) a^2-2 b x a+r^2+x^2\right]^4}\nonumber\\
    & & +\frac{1}{{48 \left(b^2 r^2+x^2\right)^3 \left(\left(b^2+1\right) r^2+x^2\right)^2 \left(\left(b^2+1\right) a^2-2 b x a+r^2+x^2\right)^3}}\nonumber\\
    & & \times\left[9 a r^4 x b^8-\left(8 r^6+9 x^2 r^4\right) b^7+6 a r^2 \left(3 x^3+7 r^2 x\right) b^6-\left(16 r^6+25 x^2 r^4+18 x^4 r^2\right) b^5\right.\nonumber\\
    & & +3 a x \left(19 r^4+12 x^2 r^2+3 x^4\right) b^4+\left(-8 r^6+8 x^2 r^4+22 x^4 r^2-9 x^6\right) b^3\nonumber\\
    & & \left. +2 a x \left(12 r^4+5 x^2 r^2-3 x^4\right) b^2+\left(39 x^6+56 r^2 x^4+24 r^4 x^2\right) b-a x^3 \left(8 r^2+15 x^2\right)\right]\nonumber\\
    & & +\frac{1}{192 \left(b^2 r^2+x^2\right)^5 \left(\left(b^2+1\right) r^2+x^2\right)^3 \left(\left(b^2+1\right) a^2-2 b x a+r^2+x^2\right)^2}\nonumber\\
    & & \times \left[45 a r^8 x b^{14}-45 r^8 x^2 b^{13}+9 a r^6 \left(20 x^3+7 r^2 x\right) b^{12}+6 r^6 \left(8 r^4-3 x^2 r^2-30 x^4\right) b^{11}\right.\nonumber\\
    & & +3 a r^4 x \left(-107 r^4+132 x^2 r^2+90 x^4\right) b^{10}+3 r^4 \left(48 r^6+x^2 r^4-72 x^4 r^2-90 x^6\right) b^9\nonumber\\
    & & +a r^2 x \left(-891 r^6+452 x^2 r^4+810 x^4 r^2+180 x^6\right) b^8\nonumber\\
    & & + 4r^2 \left(36 r^8-150 x^2 r^6-323 x^4 r^4-135 x^6 r^2-45 x^8\right) b^7\nonumber\\
    & & +a x \left(-792 r^8+916 x^2 r^6+1906 x^4 r^4+684 x^6 r^2+45 x^8\right) b^6\nonumber\\
    & & +\left(48 r^{10}-1056 x^2 r^8-2696 x^4 r^6-2038 x^6 r^4-504 x^8 r^2-45 x^{10}\right) b^5\nonumber\\
    & & +a x \left(-240 r^8+1160 x^2 r^6+2462 x^4 r^4+1172 x^6 r^2+207 x^8\right) b^4\nonumber\\
    & & - 2 x^2 \left(240 r^8+600 x^2 r^6+548 x^4 r^4+214 x^6 r^2+81 x^8\right) b^3\nonumber\\
    & & +a x^3 \left(480 r^6+1048 x^2 r^4+532 x^4 r^2+39 x^6\right) b^2\nonumber\\
    & & \left.+x^4 \left(240 r^6+768 x^2 r^4+856 x^4 r^2+363 x^6\right) b-a x^5 \left(48 r^4+136 x^2 r^2+123 x^4\right)\right]\nonumber\\
    & & +\frac{1}{128 \left(b^2 r^2+x^2\right)^7 \left(\left(b^2+1\right) r^2+x^2\right)^4 \left(\left(b^2+1\right) a^2-2 b x a+r^2+x^2\right)}\nonumber\\
    & & \times \left[45 a r^{12} x b^{20}-45 r^{12} x^2 b^{19}+54 a r^{10} x \left(2 r^2+5 x^2\right) b^{18}-9 r^{10} x^2 \left(7 r^2+30 x^2\right) b^{17}\right.\nonumber\\
    & & +9 a r^8 x \left(14 r^4+88 x^2 r^2+75 x^4\right) b^{16}-r^8 \left(64 r^6+63 x^2 r^4+522 x^4 r^2+675 x^6\right) b^{15}\nonumber\\
    & & +4 a r^6 x \left(193 r^6+147 x^2 r^4+585 x^4 r^2+225 x^6\right) b^{14}\nonumber\\
    & & -r^6 \left(256 r^8-379 x^2 r^6+66 x^4 r^4+1665 x^6 r^2+900 x^8\right) b^{13}\nonumber\\
    & & +a r^4 x \left(2413 r^8-2080 x^2 r^6+1722 x^4 r^4+3600 x^6 r^2+675 x^8\right) b^{12}\nonumber\\
    & & -r^4 \left(384 r^{10}-2904 x^2 r^8-4898 x^4 r^6+57 x^6 r^4+2700 x^8 r^2+675 x^{10}\right) b^{11}\nonumber\\
    & & +2 a r^2 x \left(1596 r^{10}-4069 x^2 r^8-3742 x^4 r^6+1764 x^6 r^4+1530 x^8 r^2+135 x^{10}\right) b^{10}\nonumber\\
    & & +\left(-256 r^{14}+5808 x^2 r^{12}+12392 x^4 r^{10}+6837 x^6 r^8-828 x^8 r^6-2385 x^{10} r^4-270 x^{12} r^2\right) b^9\nonumber\\
    & & +a x \left(1936 r^{12}-12264 x^2 r^{10}-18405 x^4 r^8-2016 x^6 r^6+4242 x^8 r^4+1368 x^{10} r^2+45 x^{12}\right) b^8\nonumber\\
    & & -\left(64 r^{14}-4672 x^2 r^{12}-8576 x^4 r^{10}-1648 x^6 r^8+3492 x^8 r^6+1857 x^{10} r^4+1098 x^{12} r^2+45 x^{14}\right) b^7\nonumber\\
    & & +4 a x \left(112 r^{12}-2128 x^2 r^{10}-3444 x^4 r^8-27 x^6 r^6+1619 x^8 r^4+651 x^{10} r^2+63 x^{12}\right) b^6\nonumber\\
    & & -x^2 \left(-1344 r^{12}+1344 x^2 r^{10}+13216 x^4 r^8+18864 x^6 r^6+9547 x^8 r^4+1506 x^{10} r^2+207 x^{12}\right) b^5\nonumber\\
    & & +a x^3 \left(-2240 r^{10}-1568 x^2 r^8+7344 x^4 r^6+10195 x^6 r^4+4096 x^8 r^2+630 x^{10}\right) b^4\nonumber\\
    & & -x^4 \left(2240 r^{10}+8512 x^2 r^8+12672 x^4 r^6+9160 x^6 r^4+3038 x^8 r^2+423 x^{10}\right) b^3\nonumber\\
    & & +2 a x^5 \left(672 r^8+2272 x^2 r^6+2668 x^4 r^4+1131 x^6 r^2+118 x^8\right) b^2\nonumber\\
    & & +x^6 \left(448 r^8+1856 x^2 r^6+2928 x^4 r^4+2120 x^6 r^2+635 x^8\right) b\nonumber\\
    & & \left.-a x^7 \left(64 r^6+240 x^2 r^4+328 x^4 r^2+187 x^6\right)\right]\nonumber\\
    & & + \frac{x^9-36 b^2 r^2 x^7+126 b^4 r^4 x^5-84 b^6 r^6 x^3+9 b^8 r^8 x}{r \left(b^2 r^2+x^2\right)^9}\arctan\left(\frac{a}{r}\right)\nonumber\\
    & & +\frac{1}{128 \left(b^2 r^2+x^2\right)^9 \left(\left(b^2+1\right) r^2+x^2\right)^{9/2}}\left[ x \left(45 r^{16} b^{24}+36 r^{14} \left(3 r^2+10 x^2\right) b^{22}\right.\right.\nonumber\\
    & & +126 r^{12} \left(r^4+8 x^2 r^2+10 x^4\right) b^{20} -84 r^{10} \left(3 r^6-10 x^2 r^4-48 x^4 r^2-30 x^6\right) b^{18}\nonumber\\
    & & -63 r^8 \left(45 r^8-24 x^2 r^6-48 x^4 r^4-144 x^6 r^2-50 x^8\right) b^{16}\nonumber\\
    & & -2520 r^6 \left(3 r^{10}-6 x^2 r^8-3 x^4 r^6-3 x^6 r^4-5 x^8 r^2-x^{10}\right) b^{14}\nonumber\\
    & & -84 r^4 \left(108 r^{12}-570 x^2 r^{10}-585 x^4 r^8-90 x^6 r^6-155 x^8 r^4-132 x^{10} r^2-15 x^{12}\right) b^{12}\nonumber\\
    & & -72 \left(72 r^{16}-924 x^2 r^{14}-1155 x^4 r^{12}+70 x^8 r^8-203 x^{10} r^6-84 x^{12} r^4-5 x^{14} r^2\right) b^{10}\nonumber\\
    & & -9 \left(128 r^{16}-4800 x^2 r^{14}-3696 x^4 r^{12}+9240 x^6 r^{10}+10010 x^8 r^8+1288 x^{10} r^6\right.\nonumber\\
    & & \left.-1120 x^{12} r^4-208 x^{14} r^2-5 x^{16}\right) b^8\nonumber\\
    & & +84 x^2 \left(128 r^{14}-288 x^2 r^{12}-1584 x^4 r^{10}-1870 x^6 r^8-780 x^8 r^6-42 x^{10} r^4+46 x^{12} r^2+3 x^{14}\right) b^6\nonumber\\
    & & +18 x^4 \left(-896 r^{12}-2880 x^2 r^{10}-3080 x^4 r^8-980 x^6 r^6+210 x^8 r^4+140 x^{10} r^2+35 x^{12}\right) b^4\nonumber\\
    & & +36 x^6 \left(128 r^{10}+560 x^2 r^8+952 x^4 r^6+770 x^6 r^4+280 x^8 r^2+35 x^{10}\right) b^2\nonumber\\
    & & \left.\left.-x^8 \left(128 r^8+576 x^2 r^6+1008 x^4 r^4+840 x^6 r^2+315 x^8\right)\right) \arctan \left(\frac{a b^2-x b+a}{\sqrt{\left(b^2+1\right) r^2+x^2}}\right) \right]\nonumber\\
    & & -\frac{r^8 b^9-36 r^6 x^2 b^7+126 r^4 x^4 b^5-84 r^2 x^6 b^3+9 x^8 b}{2 \left(b^2 r^2+x^2\right)^9}\ln \frac{\left(b^2+1\right) a^2-2 b x a+r^2+x^2}{a^2+r^2}
\end{eqnarray}

\subsection{Integral \#2-\#5}

The second to fifth integrals in Eq.~(\ref{eq:Hfunc}) can be computed using the same process described in the previous section for obtaining the integral \#1.


\subsection{Integral \#6}

To compute the sixth term on the right hand side of Eq.~(\ref{eq:Hfunc}), we need to evaluate the following integral
\begin{equation}
\label{eq:integral_six}
   \int \frac{\arctan(\frac{x-ab}{\sqrt{r^2+a^2}})}{(r^2+a^2)^{11/2}}da
\end{equation}
The trick is to integrate by parts,
$$\int fg^\prime da=fg-\int f^\prime g da$$
where $f$ and $g$ are functions of $a$, and $g^\prime \equiv \frac{dg}{da}$ and $f^\prime \equiv \frac{df}{da}$.

For the integral in Eq.~(\ref{eq:integral_six}),
\begin{equation}
    f=\arctan(\frac{x-ab}{\sqrt{r^2+a^2}})
\end{equation}
and
\begin{equation}
    g^\prime=\frac{1}{(r^2+a^2)^{11/2}}
\end{equation}
Therefore,
\begin{equation}
    f^\prime=-\dfrac{ax+br^2}{\sqrt{r^2+a^2}\left[\left(x-ab\right)^2+r^2+a^2\right]}
\end{equation}
and
\begin{equation}
    g=\dfrac{128a^9+576r^2a^7+1008r^4a^5+840r^6a^3+315r^8a}{315r^{10}\left(r^2+a^2\right)^{9/2}}
\end{equation}
Thus
\begin{eqnarray}
    & & \int \frac{\arctan(\frac{x-ab}{\sqrt{r^2+a^2}})}{(r^2+a^2)^{11/2}}da \nonumber\\
    & & = \arctan(\frac{x-ab}{\sqrt{r^2+a^2}})\dfrac{128a^9+576r^2a^7+1008r^4a^5+840r^6a^3+315r^8a}{315r^{10}\left(r^2+a^2\right)^{9/2}}\nonumber\\
    & & - \int \frac{a\left(315r^8+840a^2r^6+1008a^4r^4+576a^6r^2+128a^8\right)\left(ax+br^2\right)}{315r^{10}\left(r^2+a^2\right)^5\left[\left(x-ab\right)^2+r^2+a^2\right]}da
\end{eqnarray}


The integral in the previous equation can be broken down into the sum of various integrals, each of which is an integral of a rational function.
\begin{eqnarray}
    & & \int \frac{a\left(315r^8+840a^2r^6+1008a^4r^4+576a^6r^2+128a^8\right)\left(ax+br^2\right)}{315r^{10}\left(r^2+a^2\right)^5\left[\left(x-ab\right)^2+r^2+a^2\right]}da = \nonumber\\
    & & \int \frac{a^2x}{r^2\left(r^2+a^2\right)^5\left[\left(x-ab\right)^2+r^2+a^2\right]}da + \int \frac{8a^4x}{3r^4\left(r^2+a^2\right)^5\left[\left(x-ab\right)^2+r^2+a^2\right]}da \nonumber\\
    & & + \int \frac{16a^6x}{5r^6\left(r^2+a^2\right)^5\left[\left(x-ab\right)^2+r^2+a^2\right]} da + \int \dfrac{64a^8x}{35r^8\left(r^2+a^2\right)^5\left[\left(x-ab\right)^2+r^2+a^2\right]} da \nonumber \\
    & & + \int \frac{128a^{10}x}{315r^{10}\left(r^2+a^2\right)^5\left[\left(x-ab\right)^2+r^2+a^2\right]} da + \int \frac{8a^3b}{3r^2\left(r^2+a^2\right)^5\left[\left(x-ab\right)^2+r^2+a^2\right]}da \nonumber \\
    & & + \int\frac{16a^5b}{5r^4\left(r^2+a^2\right)^5\left[\left(x-ab\right)^2+r^2+a^2\right]}da + \int \frac{64a^7b}{35r^6\left(r^2+a^2\right)^5\left[\left(x-ab\right)^2+r^2+a^2\right]}da \nonumber\\
    & & + \int \frac{128a^9b}{315r^8\left(r^2+a^2\right)^5\left[\left(x-ab\right)^2+r^2+a^2\right]}da + \int \frac{ab}{\left(r^2+a^2\right)^5\left[\left(x-ab\right)^2+r^2+a^2\right]}da
\end{eqnarray}



Each integral above can be evaluated using Ostrogradski's method. Here we use the first term as an example. The integral to compute is
\begin{equation}
    \int \frac{a^2x}{r^2\left(r^2+a^2\right)^5\left[\left(x-ab\right)^2+r^2+a^2\right]}da
\end{equation}
Setting
\begin{equation}
    P(a)=a^2x
\end{equation}
and
\begin{equation}
    Q(a)=(r^2+a^2)^5\left[(x-ab)^2+r^2+a^2\right]
\end{equation}
then
\begin{equation}
    Q^\prime(a)=10a(a^2+r^2)^4\left[(x-ab)^2+r^2+a^2\right]+2(a^2+r^2)^5\left[(b^2+1)a-bx\right]
\end{equation}
The greatest common divisor of $Q(a)$ and $Q^\prime(a)$ is
\begin{equation}
    Q_{1}(a)=(a^2+r^2)^4 
\end{equation}
and dividing $Q(a)$ by $Q_{1}(a)$ yields
\begin{equation}
    Q_{2}(a)=(r^2+a^2)\left[(x-ab)^2+r^2+a^2\right]
\end{equation}
Therefore,
\begin{eqnarray}
\label{eq:Ostro_transform_2}
    & & \int \frac{a^2x}{r^2\left(r^2+a^2\right)^5\left[\left(x-ab\right)^2+r^2+a^2\right]}da  \nonumber\\
    & &  = \frac{1}{r^2}\left[\frac{\mathcal{A}a^7+\mathcal{B}a^6+\mathcal{C}a^5+\mathcal{D}a^4+\mathcal{E}a^3+\mathcal{F}a^2+\mathcal{G}a+\mathcal{H}}{(a^2+r^2)^4} +\int \frac{\mathcal{R}a^3+\mathcal{S}a^2+\mathcal{T}a+\mathcal{V}}{(r^2+a^2)\left[(x-ab)^2+r^2+a^2\right]}da\right]
\end{eqnarray}
Taking derivatives with respect to $a$ on both sides of Eq.~(\ref{eq:Ostro_transform_2}) and equating the corresponding coefficients yields a linear equation that can be used to determine the capital letters in the cursive font, which are all functions of $x$, $b$, and $r$. The integral term on the right hand side of Eq.~(\ref{eq:Ostro_transform_2}) has already being evaluated. The final result for the integral in Eq.~(\ref{eq:integral_six}) is
\begin{eqnarray}
    & & \int \frac{63 \arctan\left(\frac{x-a b}{\sqrt{a^2+r^2}}\right)}{256 \left(a^2+r^2\right)^{11/2}} da\nonumber\\
    &=&\frac{1}{491520} \left\{\frac{1680 \left(b r^2-a x\right)}{r^2 \left(a^2+r^2\right)^4 \left(b^2 r^2+x^2\right)}\right.\nonumber\\
    & & -\frac{40}{r^4 \left(a^2+r^2\right)^3 \left(b^2 r^2+x^2\right)^3} \left[6 a b^2 r^2 \left(28 r^2 x+47 x^3\right)+225 a b^4 r^4 x+a \left(57 x^5-56 r^2 x^3\right)\right.\nonumber\\
    & & \left. -8 b^3 \left(9 r^4 x^2+7 r^6\right)-120 b^5 r^6+24 b \left(7 r^4 x^2+2 r^2 x^4\right)\right]\nonumber\\
    & & -\frac{3}{r^8 \left(a^2+r^2\right) \left(b^2 r^2+x^2\right)^7} \left[ 34969 a b^{12} r^{12} x + 2 a b^{10} r^{10} x \left(36748 r^2+68159 x^2\right)\right.\nonumber\\
    & & +a b^8 r^8 x \left(79352 r^2 x^2+57040 r^4+214095 x^4\right)\nonumber\\
    & & +4 a b^6 r^6 x \left(-27440 r^4 x^2-19908 r^2 x^4+3920 r^6+44225 x^6\right)\nonumber\\
    & & -a b^4 r^4 x^3 \left(137760 r^4 x^2+105488 r^2 x^4+78400 r^6-84775 x^6\right)\nonumber\\
    & & +2 a b^2 r^2 x^5 \left(15520 r^4 x^2-11076 r^2 x^4+23520 r^6+11727 x^6\right)\nonumber\\
    & & +a x^7 \left(2000 r^4 x^2-2152 r^2 x^4-2240 r^6+2833 x^6\right)\nonumber\\
    & & -11968 b^{13} r^{14}-64 b^{11} r^{12} \left(233 r^2+423 x^2\right) -64 b^9 r^{10} \left(-1217 r^2 x^2+145 r^4+35 x^4\right) \nonumber\\
    & & -64 b^7 r^8 \left(-1895 r^4 x^2-3073 r^2 x^4+35 r^6-595 x^6\right)\nonumber\\
    & & +64 b^5 r^6 x^2 \left(315 r^4 x^2+1547 r^2 x^4+735 r^6+570 x^6\right)\nonumber\\
    & & -64 b^3 r^4 x^4 \left(1715 r^4 x^2+92 r^2 x^4+1225 r^6-208 x^6\right)\nonumber\\
    & & \left.+64 b r^2 x^6 \left(10 r^4 x^2-16 r^2 x^4+245 r^6+32 x^6\right)\right]\nonumber\\
    & & -\frac{2}{r^6 \left(a^2+r^2\right)^2 \left(b^2 r^2+x^2\right)^5} \left[ 15129 a b^8 r^8 x + 12 a b^6 r^6 \left(1810 r^2 x+3233 x^3\right)\right.\nonumber\\
    & & +2 a b^4 r^4 \left(4100 r^2 x^3+4200 r^4 x+17007 x^5\right)-4 a b^2 r^2 x^3 \left(3790 r^2 x^2+4200 r^4-3039 x^4\right)\nonumber\\
    & & +a x^5 \left(-1640 r^2 x^2+1680 r^4+1809 x^4\right)-5904 b^9 r^{10}-96 b^7 r^8 \left(55 r^2+81 x^2\right)\nonumber\\
    & & -48 b^5 r^6 \left(-460 r^2 x^2+35 r^4-77 x^4\right)+96 b^3 \left(175 r^8 x^2+275 r^6 x^4+74 r^4 x^6\right)\nonumber\\
    & & \left.-48 b \left(-32 r^2 x^8+20 r^4 x^6+175 r^6 x^4\right)\right]\nonumber\\
    & & -\frac{3x}{r^9 \left(b^2 r^2+x^2\right)^9}\arctan\left(\frac{a}{r}\right) \left[ 99225 b^{16} r^{16} + 264600 b^{14} r^{14} \left(r^2+2 x^2\right)\right.\nonumber\\
    & & +8820 b^{12} r^{12} \left(50 r^2 x^2+36 r^4+141 x^4\right)+7560 b^{10} r^{10} \left(-84 r^4 x^2-49 r^2 x^4+24 r^6+228 x^6\right)\nonumber\\
    & & +126 b^8 r^8 \left(-8160 r^6 x^2-14616 r^4 x^4-9060 r^2 x^6+320 r^8+12545 x^8\right)\nonumber\\
    & & -168 b^6 r^6 x^2 \left(720 r^6 x^2+2808 r^4 x^4+4555 r^2 x^6+2240 r^8-5910 x^8\right)\nonumber\\
    & & +36 b^4 r^4 x^4 \left(27360 r^6 x^2+12684 r^4 x^4-5810 r^2 x^6+15680 r^8+11445 x^8\right)\nonumber\\
    & & -72 b^2 r^2 x^6 \left(1480 r^6 x^2-588 r^4 x^4+595 r^2 x^6+2240 r^8-1400 x^8\right)\nonumber\\
    & & \left.+x^8 \left(-2880 r^6 x^2+3024 r^4 x^4-4200 r^2 x^6+4480 r^8+11025 x^8\right)\right]\nonumber\\
    & & +\frac{384 a \left(576 a^6 r^2+1008 a^4 r^4+840 a^2 r^6+128 a^8+315 r^8\right)}{r^{10} \left(r^2+a^2\right)^{9/2}}\arctan\left(\frac{x-a b}{\sqrt{r^2+a^2}}\right)\nonumber\\
    & & +\frac{384x}{r^{10} \left(b^2 r^2+x^2\right)^9 \sqrt{\left(b^2+1\right) r^2+x^2}} \left[ 315 b^{18} r^{18} + 105 b^{16} r^{16} \left(15 r^2+23 x^2\right)\right.\nonumber\\
    & & +42 b^{14} r^{14} \left(160 r^2 x^2+75 r^4+199 x^4\right)+6 b^{12} r^{12} \left(455 r^4 x^2+1939 r^2 x^4+525 r^6+2861 x^6\right)\nonumber\\
    & & +b^{10} r^{10} \left(-7980 r^6 x^2-10836 r^4 x^4+11124 r^2 x^6+1575 r^8+23063 x^8\right)\nonumber\\
    & & +b^8 r^8 \left(-9345 r^8 x^2-18732 r^6 x^4-17460 r^4 x^6+7495 r^2 x^8+315 r^{10}+20995 x^{10}\right)\nonumber\\
    & & -2 b^6 r^6 x^2 \left(105 r^8 x^2+444 r^6 x^4+3665 r^4 x^6-2210 r^2 x^8+1470 r^{10}-6460 x^{10}\right)\nonumber\\
    & & +2 b^4 r^4 x^4 \left(4635 r^8 x^2+3379 r^6 x^4-195 r^4 x^6+1020 r^2 x^8+2205 r^{10}+2584 x^{10}\right)\nonumber\\
    & & +b^2 r^2 x^6 \left(-1445 r^8 x^2+52 r^6 x^4-120 r^4 x^6+544 r^2 x^8-1260 r^{10}+1216 x^{10}\right)\nonumber\\
    & & \left.+x^8 \left(-5 r^8 x^2+8 r^6 x^4-16 r^4 x^6+64 r^2 x^8+35 r^{10}+128 x^{10}\right)\right]\times\nonumber\\
    & & \arctan\left(\frac{a b^2+a-b x}{\sqrt{\left(b^2+1\right) r^2+x^2}}\right)\nonumber\\
    & & +\frac{192}{\left(b^2 r^2+x^2\right)^9} \left[ 315 b^{17} r^8 + 420 b^{15} \left(3 r^6 x^2+r^8\right)+126 b^{13} \left(-20 r^6 x^2+15 r^4 x^4+3 r^8\right)\right. \nonumber\\
    & & +36 b^{11} \left(-168 r^6 x^2-280 r^4 x^4+35 r^2 x^6+5 r^8\right)\nonumber\\
    & & +b^9 \left(-4680 r^6 x^2-5292 r^4 x^4-10920 r^2 x^6+35 r^8+315 x^8\right)\nonumber\\
    & & -36 b^7 \left(-252 r^2 x^6-180 r^4 x^4+35 r^6 x^2+105 x^8\right)+126 b^5 \left(60 r^2 x^6+35 r^4 x^4+63 x^8\right)\nonumber\\
    & & \left.\left.-420 b^3 \left(7 r^2 x^6+9 x^8\right)+315 b x^8 \right] \ln \frac{a^2 \left(b^2+1\right)-2 a b x+r^2+x^2}{r^2+a^2} \right\}
\end{eqnarray}

Combining all the results above, we get the final expression of $H(x,a,r)$, which can be used to express the integrated repulsive potential (i.e., the integration of the $1/r^{12}$ term) between two thin rods.
\begin{eqnarray}
\label{eq:func_H_xar}
     & & H(x,a,r) = \frac{1}{3840  r^8(a^2+r^2)^4 \left((b^2+1)r^2+x^2\right)^4 \left(a^2(b^2+1)-2 a b x+r^2+x^2\right)^4}\times \nonumber \\
     & &  \left\{ 3 (b^2+1 )^4 x \left[ (187 (b^2+1 )^3 r^6+328 (b^2+1 )^2 x^2 r^4+240 (b^2+1 ) x^4 r^2+64 x^6 \right] a^{15} \right. \nonumber \\
     & & -3 b (b^2+1 )^3 \left[ 64 (b^2+1 )^4 r^8+1565 (b^2+1 )^3 x^2 r^6+2680 (b^2+1 )^2 x^4 r^4 +1936 (b^2+1 ) x^6 r^2+512 x^8 \right] a^{14} \nonumber \\
     & & +(b^2+1 )^2 x \left[3 (b^2+1 )^4 (1196 b^2+1455 ) r^8+(b^2+1 )^3 (21093 b^2+9857 ) x^2 r^6\right.\nonumber\\
     & & \left. +8 (b^2+1 )^2 (3951 b^2+1189 ) x^4 r^4+48 (b^2+1) (443 b^2+91)  x^6 r^2 +768 (7 b^2+1 ) x^8 \right] a^{13} \nonumber \\
     & & -b (b^2+1 ) \left[ 48 (b^2+1 )^5 (15 b^2+29 ) r^{10}+3 (b^2+1 )^4 (7540 b^2+10883 ) x^2 r^8 +(b^2+1 )^3 (67635 b^2+68581 ) x^4 r^6 \right.\nonumber \\
     & & \left. +8 (b^2+1 )^2 (10209 b^2+7937 ) x^6 r^4 +192 (b^2+1) (247 b^2+145) x^8 r^2+1536 (7 b^2+3 ) x^{10} \right] a^{12}\nonumber \\
     & & +x \left[3 (b^2+1)^5 (2802 b^4+7856 b^2+4923) r^{10}+2 (b^2+1)^4 (39954 b^4+70537 b^2+20405) x^2 r^8 \right. \nonumber \\
     & & +(b^2+1)^3 (163731 b^4+242506 b^2+48663) x^4 r^6 +8 (b^2+1)^2 (19569 b^4+24994 b^2+3793) x^6 r^4 \nonumber \\
     & & \left. +96 (b^2+1) (763 b^4+822 b^2+99) x^8 r^2 +384 (35 b^4+30 b^2+3) x^{10} \right] a^{11} \nonumber \\
     & & -b \left[8 (b^2+1 )^5 (123 b^4+562 b^2+543 ) r^{12} +(b^2+1 )^4 (42618 b^4+132008 b^2+96915 ) x^2 r^{10} \right. \nonumber \\
     & & +2 (b^2+1 )^3 (96486 b^4+216701 b^2+119915 ) x^4 r^8 +(b^2+1 )^2 (301317 b^4+585674 b^2+267109 ) x^6 r^6\nonumber \\
     & & \left. +16 (b^2+1) (14457 b^4+25234 b^2+9681) x^8 r^4 +96 (875 b^4+1390 b^2+459 ) x^{10} r^2+1536 (7 b^2+3 ) x^{12} \right] a^{10} \nonumber \\
     & & + x \left[ 768 (7 b^2+1 ) x^{12}+96 (763 b^4+822 b^2+99 ) r^2 x^{10} +16 (16485 b^6+36665 b^4+23115 b^2+2727 ) r^4 x^8 \right. \nonumber \\
     & & +(b^2+1 ) (428055 b^6+1048741 b^4+734629 b^2+101399) r^6 x^6\nonumber \\
     & & +(b^2+1 )^2 (351564 b^6+970913 b^4+767622 b^2+131185 ) r^8 x^4\nonumber \\
     & & +(b^2+1 )^3 (127374 b^6+424378 b^4+398879 b^2+92579 ) r^{10} x^2 \nonumber \\
     & & \left. +(b^2+1 )^4 (9132 b^6+45706 b^4+65580 b^2+28317 ) r^{12} \right] a^9 \nonumber \\
     & & -b \left[(b^2+1 )^4 (561 b^6+5153 b^4+11767 b^2+7575 ) r^{14}\right. \nonumber \\
     & & +(b^2+1 )^3 (38616 b^6+201070 b^4+318792 b^2+158913 ) x^2 r^{12} \nonumber \\
     & & +(b^2+1 )^2 (256584 b^6+947360 b^4+1136303 b^2+449877 ) x^4 r^{10}\nonumber \\
     & & +(b^2+1 ) (502296 b^6+1629001 b^4+1704494 b^2+580869) x^6 r^8\nonumber \\
     & & +2 (239793 b^6+710497 b^4+674295 b^2+203991 ) x^8 r^6\nonumber \\
     & & \left. +16 (14457 b^4+25234 b^2+9681 ) x^{10} r^4+192 (247 b^2+145 ) x^{12} r^2 +1536 x^{14} \right] a^8 \nonumber \\
     & & +x \left[ 192 x^{14}+48 (443 b^2+91 ) r^2 x^{12}+8 (19569 b^4+24994 b^2+3793 ) r^4 x^{10} \right.\nonumber \\
     & & +(428055 b^6+1048741 b^4+734629 b^2+101399 ) r^6 x^8\nonumber \\
     & & +8 (b^2+1 ) (70755 b^6+192841 b^4+151689 b^2+23715) r^8 x^6\nonumber \\
     & & +4 (b^2+1 )^2 (94407 b^6+299896 b^4+273347 b^2+51922 ) r^{10} x^4\nonumber \\
     & & +4 (b^2+1 )^3 (24564 b^6+106463 b^4+122081 b^2+31654 ) r^{12} x^2\nonumber \\
     & & \left. +2 (b^2+1 )^4 (2244 b^6+17582 b^4+33093 b^2+16761 ) r^{14} \right] a^7 \nonumber \\
     & & -b r^2 \left[ (b^2+1 )^4 (1929 b^4+8150 b^2+7701 ) r^{14} +2 (b^2+1 )^3 (7854 b^6+60694 b^4+124841 b^2+76911 ) x^2 r^{12} \right. \nonumber \\
     & & +2 (b^2+1 )^2 (83496 b^6+378821 b^4+531204 b^2+244219 ) x^4 r^{10}\nonumber \\
     & & +4 (b^2+1 ) (106743 b^6+393911 b^4+462833 b^2+178625) x^6 r^8\nonumber \\
     & & +(502296 b^6+1629001 b^4+1704494 b^2+580869 ) x^8 r^6\nonumber \\
     & & \left. +(301317 b^4+585674 b^2+267109 ) x^{10} r^4+8 (10209 b^2+7937 ) x^{12} r^2+5808 x^{14} \right] a^6 \nonumber \\
     & & +r^2 x \left[6 (b^2+1 )^4 (1929 b^4+6248 b^2+4110 ) r^{14} \right. \nonumber \\
     & & +2 (b^2+1 )^3 (15708 b^6+107810 b^4+167127 b^2+52897 ) x^2 r^{12} \nonumber \\
     & & +4 (b^2+1 )^2 (49602 b^6+194669 b^4+215756 b^2+49377 ) x^4 r^{10}\nonumber \\
     & & +4 (b^2+1 ) (94407 b^6+299896 b^4+273347 b^2+51922) x^6 r^8 \nonumber \\
     & & +(351564 b^6+970913 b^4+767622 b^2+131185 ) x^8 r^6\nonumber \\
     & & \left. +(163731 b^4+242506 b^2+48663 ) x^{10} r^4+8 (3951 b^2+1189 ) x^{12} r^2+720 x^{14} \right] a^5 \nonumber \\
     & & -b r^4 \left[ 15 (b^2+1 )^4 (153 b^2+283 ) r^{14}+15 (b^2+1 )^3 (1929 b^4+6650 b^2+5631 ) x^2 r^{12}\right.\nonumber \\
     & & +10 (b^2+1 )^2 (3927 b^6+28439 b^4+51979 b^2+29807 ) x^4 r^{10}\nonumber \\
     & & +2 (b^2+1 ) (83496 b^6+378821 b^4+531204 b^2+244219) x^6 r^8\nonumber \\
     & & +(256584 b^6+947360 b^4+1136303 b^2+449877 ) x^8 r^6\nonumber \\
     & & \left. +2 (96486 b^4+216701 b^2+119915 ) x^{10} r^4+(67635 b^2+68581 ) x^{12} r^2+8040 x^{14} \right] a^4 \nonumber \\
     & & +r^4 x \left[ 180 (b^2+1 )^4 (51 b^2+58 ) r^{14}+20 (b^2+1 )^3 (1929 b^4+5498 b^2+2513 ) x^2 r^{12}\right.\nonumber \\
     & & +2 (b^2+1 )^2 (15708 b^6+107810 b^4+167127 b^2+52897 ) x^4 r^{10}\nonumber \\
     & & +4 (b^2+1 ) (24564 b^6+106463 b^4+122081 b^2+31654) x^6 r^8\nonumber \\
     & & +(127374 b^6+424378 b^4+398879 b^2+92579 ) x^8 r^6\nonumber \\
     & & \left. +2 (39954 b^4+70537 b^2+20405 ) x^{10} r^4+(21093 b^2+9857 ) x^{12} r^2+984 x^{14} \right] a^3 \nonumber \\
     & & -b r^6 \left[ 975 (b^2+1 )^4 r^{14}+30 (b^2+1 )^3 (459 b^2+719 ) x^2 r^{12} +15 (b^2+1 )^2 (1929 b^4+6650 b^2+5631 ) x^4 r^{10} \right. \nonumber \\
     & & + 2 (b^2+1 ) (7854 b^6+60694 b^4+124841 b^2+76911) x^6 r^8\nonumber \\
     & & +(38616 b^6+201070 b^4+318792 b^2+158913 ) x^8 r^6\nonumber \\
     & & \left. +(42618 b^4+132008 b^2+96915 ) x^{10} r^4+3 (7540 b^2+10883 ) x^{12} r^2+4695 x^{14} \right] a^2 \nonumber \\
     & & +r^6 x \left[ 1950 (b^2+1 )^4 r^{14}+180 (b^2+1 )^3 (51 b^2+58 ) x^2 r^{12} +6 (b^2+1 )^2 (1929 b^4+6248 b^2+4110 ) x^4 r^{10} \right. \nonumber \\
     & & +2 (b^2+1) (2244 b^6+17582 b^4+33093 b^2+16761) x^6 r^8 +(9132 b^6+45706 b^4+65580 b^2+28317 ) x^8 r^6 \nonumber \\
     & & \left. +3 (2802 b^4+7856 b^2+4923 ) x^{10} r^4 +3 (1196 b^2+1455 ) x^{12} r^2+561 x^{14} \right] a \nonumber \\
     & & -br^8 x^2 \left[ 975 (b^2+1 )^3 r^{12}+15 (b^2+1 )^2 (153 b^2+283 ) x^2 r^{10} +(b^2+1 ) (1929 b^4+8150 b^2+7701) x^4 r^8 \right. \nonumber \\
     & &\left. \left. +(561 b^6+5153 b^4+11767 b^2+7575 ) x^6 r^6+8 (123 b^4+562 b^2+543 ) x^8 r^4 +48 (15 b^2+29 ) x^{10} r^2+192 x^{12}\right] \right\}  \nonumber \\
     & & +\frac{ 315 \left(b^2+1\right)^4 r^8 x + 840 \left(b^2+1\right)^3 r^6 x^3+1008 \left(b^2+1\right)^2 r^4 x^5+576 \left(b^2+1\right) r^2 x^7 +128 x^9}{1280 r^{10} \left(\left(b^2+1\right) r^2+x^2\right)^{9/2}}\nonumber \\
     & & \times \arctan \left(\frac{a b^2+a-b x}{\sqrt{\left(b^2+1\right) r^2+x^2}}\right) \nonumber \\
     & & +\frac{315 r^8a + 840 r^6 a^3 + 1008 r^4 a^5 + 576 r^2 a^7 + 128 a^9}{1280 r^{10} \left(a^2+r^2\right)^{9/2}}\times \arctan \left(\frac{x-a b}{\sqrt{a^2+r^2}}\right)
\end{eqnarray}

\section{Integrated Attractive Potential}

The expressions for the integrated attractive potential (i.e., the integration of the $1/r^6$ term) can be derived in similar fashions. The final results are included in the main text.






\section{Forces from Rod-Rod Interactions}

Using the affine coordinates defined in Fig.~1 of the main text, we can express the displacement, $d\vb{s}$, of the rod 1 as
\begin{equation}
    \label{eq:displacement}
    d\vb{s}=dx \vb{n}_x- d\rho \vb{n}_{\rho} - dr \vb{n}_z
\end{equation}
Note that in Fig.~1 of the main text, the center of rod 1 is located at $(x,0,0)$ (for simplicity, we denote $x_0$ as $x$ in this section) and the center of rod 2 is located at $(0,\rho,r)$ (we denote $\rho_0$ as $\rho$ in this section). The configuration of the two-rod system is given by $(x,\rho,r)$ and a configuration change $(dr,d\rho,dr)$ corresponds to a displacement of the rod 1 given in Eq.~(\ref{eq:displacement}).


For a scalar function $U(x,\rho, r)$ describing the potential between the two rods, its total derivative can be written as
\begin{equation}
    dU= \nabla{U} \cdot d\vb{s}
\end{equation}
where $\nabla{U}$ is the gradient in the affine coordinate system. Meanwhile, the total derivative of $U$ can also be written as
\begin{equation}
    dU=\frac{\partial U}{\partial x}dx+\frac{\partial U}{\partial \rho}d\rho+\frac{\partial U}{\partial r}dr
\end{equation}
Combining those two equations, we have
\begin{equation}
    \label{eq:total_derivative}
    \nabla{U} \cdot d\vb{s}=\frac{\partial U}{\partial x}dx+\frac{\partial U}{\partial \rho}d\rho+\frac{\partial U}{\partial r}dr
\end{equation}
We can write the gradient as
\begin{equation}
    \label{eq:nabla}
    \nabla{U}=\alpha \vb{n}_x+ \beta \vb{n}_{\rho}+\gamma \vb{n}_z
\end{equation}
Plugging Eqs.~(\ref{eq:displacement}) and (\ref{eq:nabla}) into Eq.~(\ref{eq:total_derivative}), the left hand side of Eq.~(\ref{eq:total_derivative}) becomes
\begin{equation}
    (\alpha \vb{n}_x+ \beta \vb{n}_{\rho}+\gamma \vb{n}_z)\cdot(dx \vb{n}_x- d\rho \vb{n}_{\rho} - dr \vb{n}_z )
\end{equation}
For the unit basis of the affine coordinate system,
\begin{eqnarray}
    & & \vb{n}_x\cdot \vb{n}_x=1 \quad\quad \vb{n}_{\rho}\cdot \vb{n}_{\rho}=1 \quad\quad \vb{n}_z\cdot \vb{n}_z=1\nonumber\\
    & & \vb{n}_x\cdot \vb{n}_z=0 \quad\quad \vb{n}_{\rho}\cdot \vb{n}_z=0 \quad\quad \vb{n}_x\cdot \vb{n}_{\rho}=\cos\theta
\end{eqnarray}
Thus Eq.~(\ref{eq:total_derivative}) yields
\begin{equation}
    (\alpha +\beta \cos\theta)dx-(\alpha \cos\theta+\beta)d\rho - \gamma dr=\frac{\partial U}{\partial x}dx+\frac{\partial U}{\partial \rho}d\rho+\frac{\partial U}{\partial r}dr
\end{equation}
We arrive at the following set of equations,
\begin{eqnarray}
    \frac{\partial U}{\partial x} &=& \alpha + \beta \cos\theta \nonumber\\
    \frac{\partial U}{\partial \rho} &=& -\alpha \cos\theta - \beta \nonumber\\
    \frac{ \partial U}{\partial r} &=& - \gamma
\end{eqnarray}
We can then solve for $\alpha$, $\beta$, and $\gamma$ and the results are
\begin{eqnarray}
    \alpha &=& \frac{1}{\sin^2{\theta}}\left(\frac{\partial U}{\partial x}+ \cos{\theta} \frac{\partial U}{\partial \rho} \right) \nonumber\\
    \beta &=& -\frac{1}{\sin^2{\theta}}\left(\frac{\partial U}{\partial  \rho}+ \cos{\theta} \frac{\partial U}{\partial x}\right)\nonumber\\
    \gamma &=& -\frac{\partial U}{\partial r}
\end{eqnarray}
The force on the rod 1 is
\begin{eqnarray}
    \vb{F}_1 &=&-\nabla{U}\nonumber\\
    &=& -\left(\alpha \vb{n}_x+ \beta \vb{n}_{\rho}+\gamma \vb{n}_z\right)\nonumber\\
    &=& -\frac{1}{\sin^2{\theta}}\left(\frac{\partial U}{\partial x}+ \cos{\theta} \frac{\partial U}{\partial \rho}\right)\vb{n}_x + \frac{1}{\sin^2{\theta}}\left(\frac{\partial U}{\partial  \rho}+ \cos{\theta} \frac{\partial U}{\partial x}\right) \vb{n}_{\rho} + \frac{\partial U}{\partial r}\vb{n}_z
\end{eqnarray}
The force on the rod 2 is $\vb{F}_2 = -\vb{F}_1$.


\section{Torques from Rod-Rod Interactions}

This section follows the affine coordinates and the notation in Fig.~1 of the main text. The center of rod 1 is located at $(x_0, 0, 0)$ and the center of rod 2 is located at $(0, \rho_0, r)$. If we denote the force on a volume element $dx$ of rod 1 from its interaction with a volume element $d\rho$ of rod 2 as $\vb{f}_1$ and the corresponding potential as $u(x,\rho,r,\theta)$. It can be easily proven the following identities.
\begin{equation}
    \frac{\rho}{\sin\theta} \left( \frac{\partial u}{\partial x} + \cos \theta \frac{\partial u}{\partial \rho} \right) = \frac{\partial u}{\partial \theta} 
\end{equation}
\begin{equation}
    \frac{x}{\sin\theta} \left( \frac{\partial u}{\partial \rho} + \cos \theta \frac{\partial u}{\partial x} \right) = \frac{\partial u}{\partial \theta} 
\end{equation}
\begin{equation}
    \frac{r}{x\sin^2\theta} \left( \frac{\partial u}{\partial x} + \cos \theta \frac{\partial u}{\partial \rho} \right) = \frac{\partial u}{\partial r} 
\end{equation}
\begin{equation}
    \frac{r}{\rho\sin^2\theta} \left( \frac{\partial u}{\partial \rho} + \cos \theta \frac{\partial u}{\partial x} \right) = \frac{\partial u}{\partial r} 
\end{equation}

The torque on rod 1 has the following component along $\vb{n}_z$.
\begin{eqnarray}
    \boldsymbol{\tau}_{1z} &=& \int\int (x-x_0)\vb{n}_x\times \vb{f}_{1\rho} d x d\rho\nonumber\\
    &=&\int\int x f_{1\rho}\sin \theta dxd\rho \vb{n}_z-x_0 \vb{n}_x \times \vb{F}_{1\rho} \nonumber\\
    &=&\int\int \frac{x}{\sin\theta}\left(\frac{\partial u}{\partial \rho}+\frac{\partial u}{\partial x}\cos\theta\right)dxd\rho \vb{n}_z -x_0\vb{n}_x\times \vb{F}_{1\rho}\nonumber\\
    &=&\int\int \frac{\partial u}{\partial \theta}dxd\rho \vb{n}_z-x_0\vb{n}_x\times \vb{F}_{1\rho}\nonumber \\
    &=&\frac{\partial }{\partial \theta} \int \int u dxd\rho \vb{n}_z-x_0\vb{n}_x\times \vb{F}_{1\rho}\nonumber\\
    &=&\frac{\partial U}{\partial \theta} \vb{n}_z-x_0\vb{n}_x\times \vb{F}_{1\rho} \nonumber \\
    &=& \left(\frac{\partial U}{\partial \theta}-x_0 F_{1\rho} \sin \theta \right)\vb{n}_z \nonumber \\
    &=& \left[\frac{\partial U}{\partial \theta}-\frac{x_0}{\sin\theta} \left(\frac{\partial U}{\partial  \rho}+ \cos{\theta} \frac{\partial U}{\partial x}\right)\right]\vb{n}_z
\end{eqnarray}
Here $U \equiv U(x_0, \rho_0, r, \theta)$ is the integrated rod-rod potential.

The other component of the torque on rod 1 is
\begin{eqnarray}
    \boldsymbol{\tau}_{1y}&=& \int\int (x-x_0)\vb{n}_x \times \vb{f}_{1z}dxd\rho \nonumber \\
    &=& \int\int x\vb{n}_x \times \vb{f}_{1z}dxd\rho-x_0 \vb{n}_x\times \vb{F}_{1z}\nonumber \\
    &=& \int\int x\vb{n}_x \times \frac{\partial u}{\partial r}\vb{n}_z dxd\rho - x_0 \vb{n}_x\times \vb{F}_{1z}\nonumber \\
    &=& -\int\int x\frac{\partial u}{\partial r} dxd\rho \vb{n}_y - x_0 \vb{n}_x\times \vb{F}_{1z}\nonumber \\
    &=& -\int\int \frac{r}{\sin^2\theta} \left( \frac{\partial u}{\partial x} + \cos \theta \frac{\partial u}{\partial \rho} \right) dxd\rho \vb{n}_y -x_0 \vb{n}_x\times \vb{F}_{1z}\nonumber \\
    &=& r\int \int f_{1x}dxd\rho \vb{n}_y -x_0 \vb{n}_x\times \vb{F}_{1z}\nonumber \\
    &=& r F_{1x} \vb{n}_y -x_0 \vb{n}_x\times \vb{F}_{1z}\nonumber \\
    &=& r\vb{n}_z\times \vb{F}_{1x}-x_0 \vb{n}_x\times \vb{F}_{1z} \nonumber \\
    &=& \left(r F_{1x}+x_0 F_{1z} \right) \vb{n}_y \nonumber \\
    &=& \left[ -\frac{r}{\sin^2\theta} \left( \frac{\partial U}{\partial x} + \cos\theta \frac{\partial U}{\partial \rho} \right) + x_0 \frac{\partial U}{\partial r} \right] \vb{n}_y
\end{eqnarray}
Here $\vb{n}_y = \vb{n}_z \times \vb{n}_x$ setting a local right-handed Cartesian coordinate system $(xyz)$.

The torque on rod 2 can be derived similarly.
\begin{eqnarray}
    \boldsymbol{\tau}_{2z} &=& \int\int (\rho-\rho_0)\vb{n}_\rho\times \vb{f}_{2x} d x d\rho\nonumber\\
    &=& - \int\int \rho f_{2x}\sin \theta dxd\rho \vb{n}_z-\rho_0 \vb{n}_\rho \times \vb{F}_{2x} \nonumber\\
    &=& - \int\int \frac{\rho}{\sin\theta}\left(\frac{\partial u}{\partial x}+\frac{\partial u}{\partial \rho}\cos\theta\right)dxd\rho \vb{n}_z -\rho_0 \vb{n}_\rho \times \vb{F}_{2x}\nonumber\\
    &=& - \int\int \frac{\partial u}{\partial \theta}dxd\rho \vb{n}_z-\rho_0 \vb{n}_\rho \times \vb{F}_{2x}\nonumber \\
    &=& - \frac{\partial }{\partial \theta} \int \int u dxd\rho \vb{n}_z-\rho_0 \vb{n}_\rho \times \vb{F}_{2x}\nonumber\\
    &=& - \frac{\partial U}{\partial \theta} \vb{n}_z-\rho_0 \vb{n}_\rho \times \vb{F}_{2x} \nonumber \\
    &=& \left(-\frac{\partial U}{\partial \theta}+\rho_0 F_{2x} \sin \theta \right)\vb{n}_z \nonumber \\
    &=& \left[- \frac{\partial U}{\partial \theta}+\frac{\rho_0}{\sin\theta} \left(\frac{\partial U}{\partial  x}+ \cos{\theta} \frac{\partial U}{\partial \rho}\right)\right]\vb{n}_z
\end{eqnarray}
and
\begin{eqnarray}
    \boldsymbol{\tau}_{2t}&=& \int\int (\rho-\rho_0)\vb{n}_\rho \times \vb{f}_{2z}dxd\rho \nonumber \\
    &=& \int\int \rho\vb{n}_\rho \times \vb{f}_{2z}dxd\rho-\rho_0 \vb{n}_\rho\times \vb{F}_{2z}\nonumber \\
    &=& - \int\int \rho\vb{n}_\rho \times \frac{\partial u}{\partial r}\vb{n}_z dxd\rho - \rho_0 \vb{n}_\rho\times \vb{F}_{2z}\nonumber \\
    &=& \int\int \rho\frac{\partial u}{\partial r} dxd\rho \vb{n}_t - \rho_0 \vb{n}_\rho\times \vb{F}_{2z}\nonumber \\
    &=& \int\int \frac{r}{\sin^2\theta} \left( \frac{\partial u}{\partial \rho} + \cos \theta \frac{\partial u}{\partial x} \right) dxd\rho \vb{n}_t -\rho_0 \vb{n}_\rho\times \vb{F}_{2z}\nonumber \\
    &=& -r\int \int f_{2\rho}dxd\rho \vb{n}_t -\rho_0 \vb{n}_\rho\times \vb{F}_{2z}\nonumber \\
    &=& -r F_{2\rho} \vb{n}_t -\rho_0 \vb{n}_\rho\times \vb{F}_{2z}\nonumber \\
    &=& -r\vb{n}_z\times \vb{F}_{2\rho}-\rho_0 \vb{n}_\rho\times \vb{F}_{2z} \nonumber \\
    &=& \left(-r F_{2\rho}+\rho_0 F_{2z} \right) \vb{n}_t \nonumber \\
    &=& \left[ \frac{r}{\sin^2\theta} \left( \frac{\partial U}{\partial \rho} + \cos\theta \frac{\partial U}{\partial x} \right) - \rho_0 \frac{\partial U}{\partial r} \right] \vb{n}_t
\end{eqnarray}
Here $\vb{n}_t = \vb{n}_z \times \vb{n}_\rho$ setting another local right-handed Cartesian coordinate system $(\rho t z)$.

\section{Forces and Torques from Bead-Rod Interactions}

Using the affine coordinates defined in Fig.~2 of the main text, the point particle (i.e., the bead) is located at $(\rho, \theta)$. Therefore, the force on the bead is
\begin{eqnarray}
    \vb{F}_2 & =&  -\nabla W(\rho,\theta) \nonumber\\
    &=& -\frac{\partial W}{\partial \rho} \vb{n}_\rho
    - \frac{1}{\rho} \frac{\partial W}{\partial \theta} \vb{n}_\theta
\end{eqnarray}
Note that
\begin{equation}
    \vb{n}_\theta = -\frac{1}{\sin\theta} \vb{n}_x
    + \frac{\cos\theta}{\sin\theta} \vb{n}_\rho
\end{equation}
Therefore
\begin{equation}
    \vb{F}_2  = \frac{1}{\rho \sin\theta} \frac{\partial W}{\partial \theta} \vb{n}_x -\left( \frac{\partial W}{\partial \rho} +
     \frac{\cos\theta}{\rho\sin\theta} \frac{\partial W}{\partial \theta} \right) \vb{n}_\rho
\end{equation}
The force on the rod is $\vb{F}_1 = - \vb{F}_2$.

The interaction between the point particle at $(\rho,\theta)$ and a volume element $dx$ on the rod is
\begin{equation}
    w(x,\rho,\theta) = \frac{B}{\left( x^2 +\rho^2 -2x\rho\cos\theta\right)^6} - \frac{A}{\left( x^2 +\rho^2 -2x\rho\cos\theta\right)^3}
\end{equation}
The integration of $w(x,\rho,\theta)$ over $x$ yields the full potential $W(\rho,\theta)$. It is easy to show that
\begin{equation}
    x\left(\frac{\partial w}{\partial \rho} +
     \frac{\cos\theta}{\rho\sin\theta} \frac{\partial w}{\partial \theta} \right) = \frac{1}{\sin\theta} \frac{\partial w}{\partial \theta}
\end{equation}

The torque on the rod can be computed as
\begin{eqnarray}
    \boldsymbol{\tau}_{1z} &=& \int (x-l)\vb{n}_x\times \vb{f}_{1\rho} d x \nonumber\\
    &=& \int x \left(\frac{\partial w}{\partial \rho} +
     \frac{\cos\theta}{\rho\sin\theta} \frac{\partial w}{\partial \theta} \right)\sin\theta dx \vb{n}_z - l \vb{n}_x \times \vb{F}_{1\rho} \nonumber \\
     & =& \int \frac{\partial w}{\partial \theta} dx \vb{n}_z - l \vb{n}_x \times \vb{F}_{1\rho} \nonumber \\
     &=& \frac{\partial }{\partial \theta} \int w dx \vb{n}_z - l \vb{n}_x \times \vb{F}_{1\rho} \nonumber \\
     &=& \frac{\partial W}{\partial \theta} \vb{n}_z - l \vb{n}_x \times \vb{F}_{1\rho} \nonumber \\
     &=& \left[ \frac{\partial W}{\partial \theta} -l \left( \frac{\partial W}{\partial \rho} +
     \frac{\cos\theta}{\rho\sin\theta} \frac{\partial W}{\partial \theta} \right)\sin\theta \right]\vb{n}_z \nonumber \\
     &=& \left[ -l \sin\theta \frac{\partial W}{\partial \rho} + \left( 1- \frac{l}{\rho} \cos\theta \right) \frac{\partial W}{\partial \theta} \right]\vb{n}_z
\end{eqnarray}
Here $\vb{n}_z \equiv \frac{1}{\sin\theta} \vb{n}_x\times \vb{n}_\rho$ is the unit vector perpendicular to the $x\rho$ plane.

\section{Verification of Analytical Results}

Since our analytical results on the integrated potentials, forces, and torques involve very lengthy expressions, we compare them to the results from direct numerical integration in this section. An essentially perfect agreement is found in all cases, as expected. Such comparison thus completely verifies the correctness of the analytical results presented in the main text. Some examples of the comparison between the two are included below.


\begin{figure}[htb]
    \centering
    \includegraphics[width=1\textwidth]{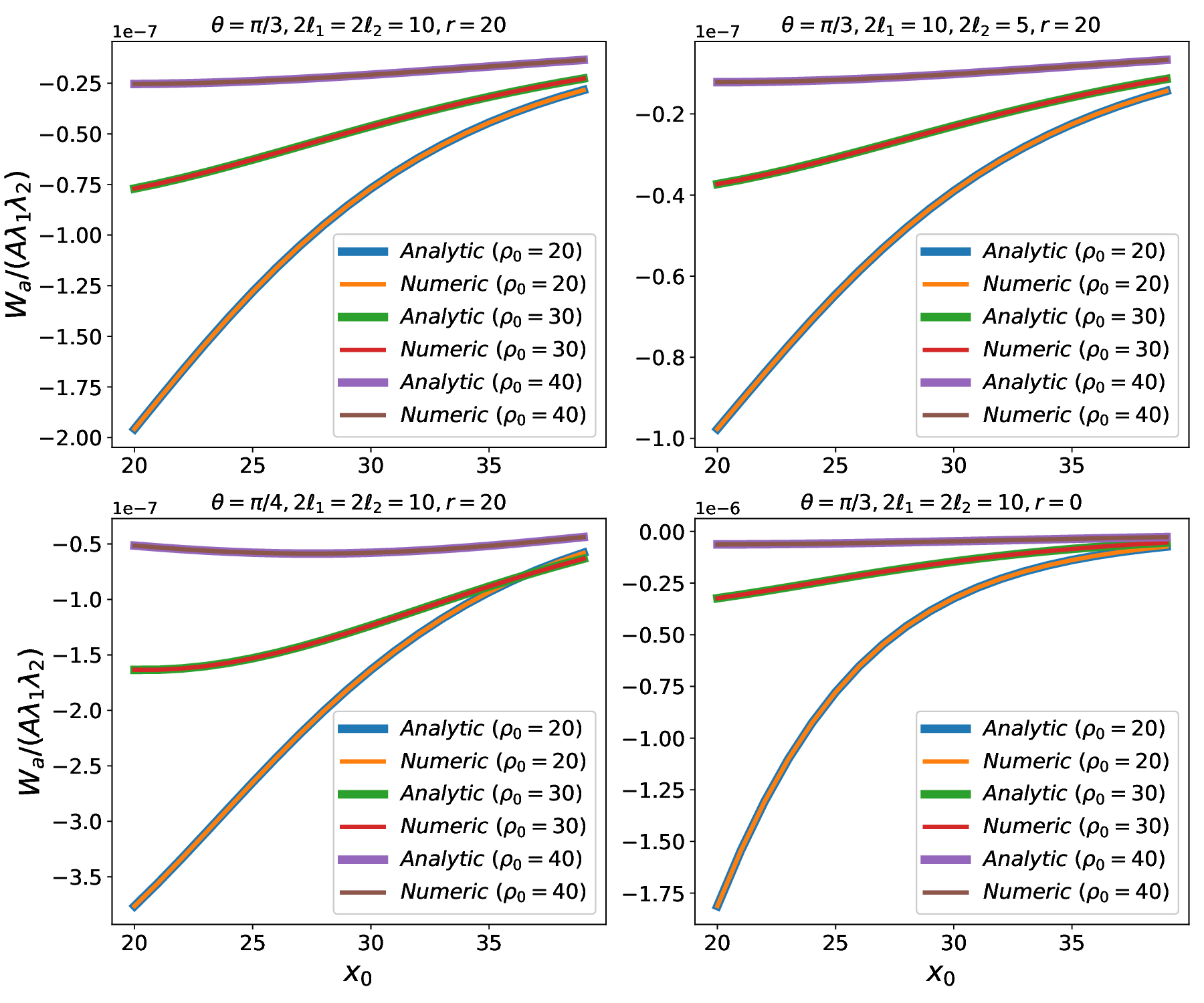}
    \caption{Comparison of analytical and numerical results for the integrated attractive potential between a pair of thin rods in various situations.}
    \label{fg:attr_comparison}
\end{figure}


\begin{figure}[htb]
    \centering
    \includegraphics[width=1\textwidth]{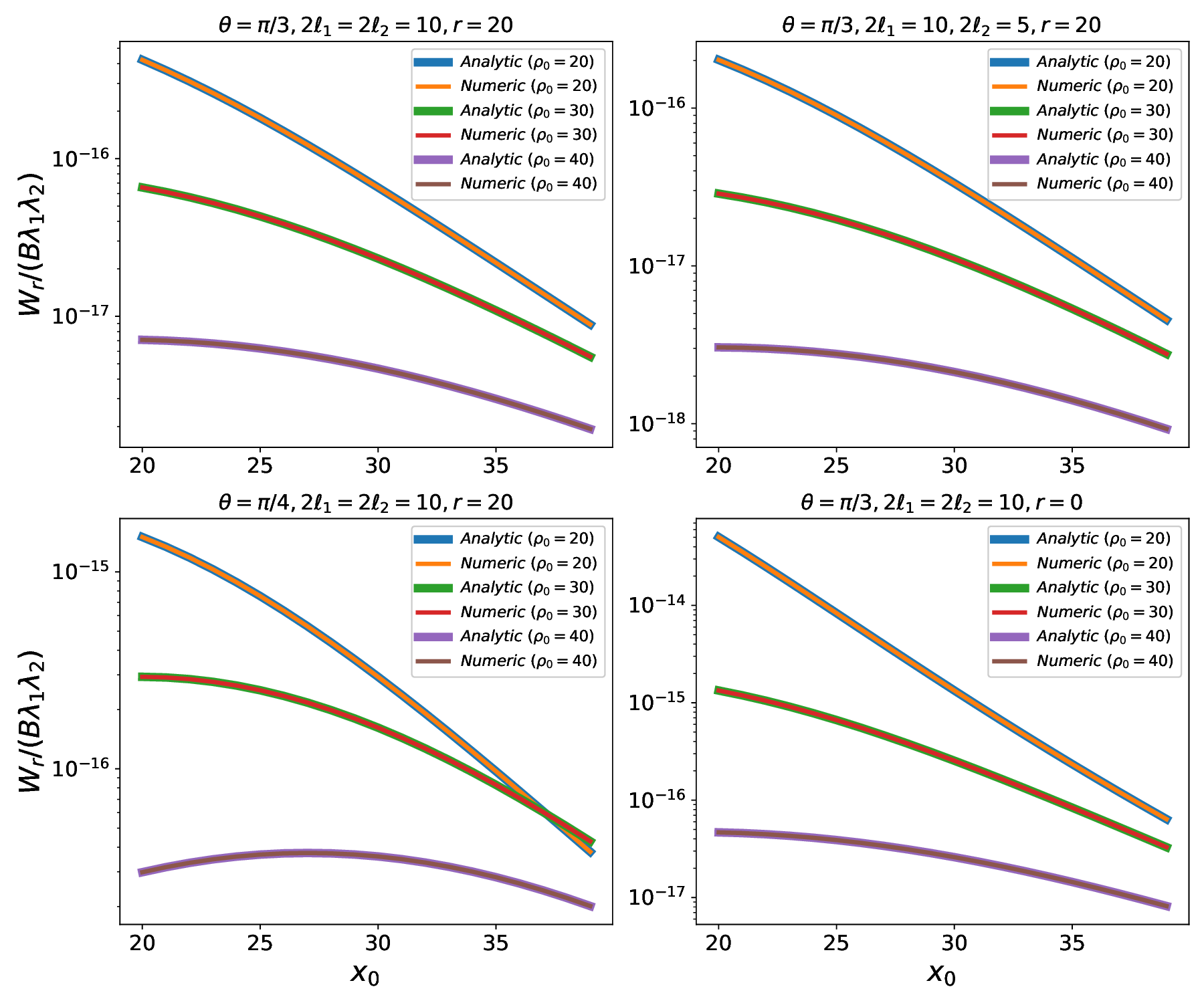}
    \caption{Comparison of analytical and numerical results for the integrated repulsive potential between a pair of thin rods in various situations.}
    \label{fg:repul_comparison}
\end{figure}



\begin{figure}[htb]
    \centering
    \includegraphics[width=1\textwidth]{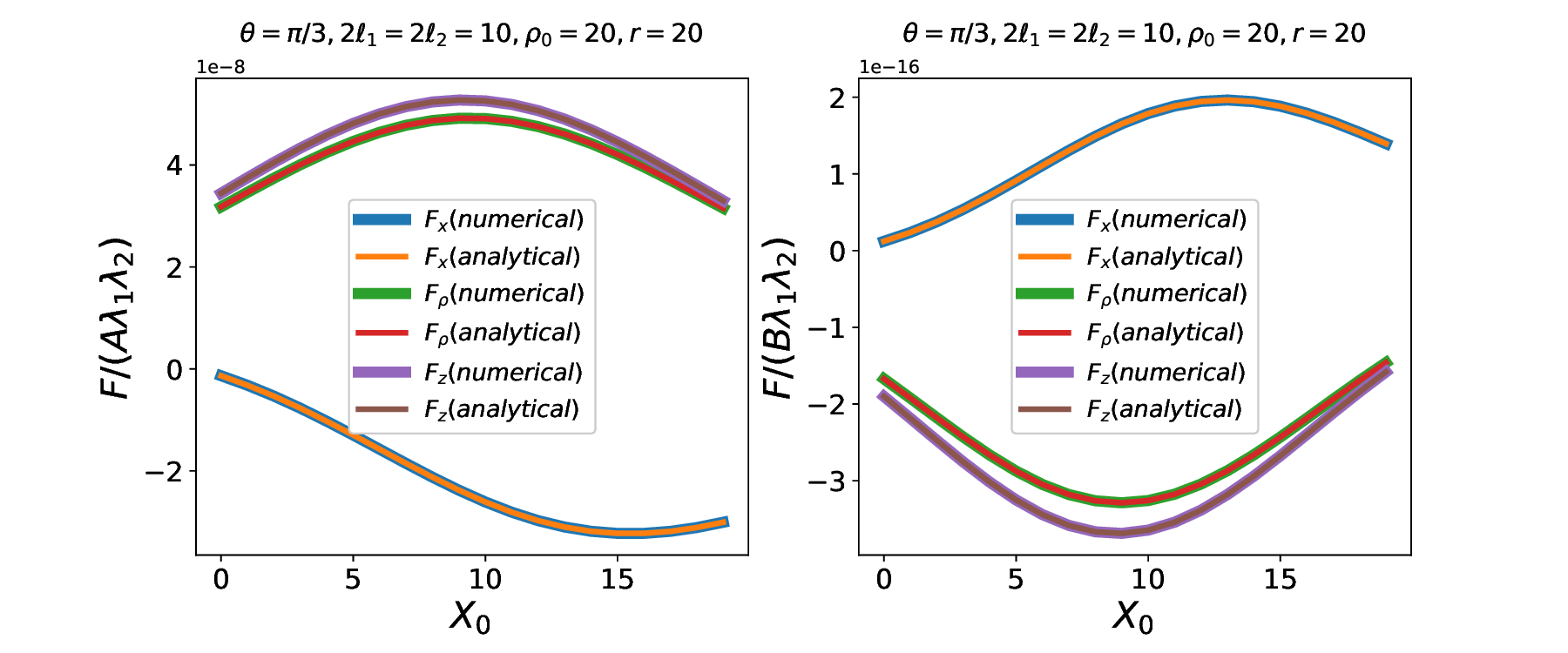}
    \caption{Comparison of analytical and numerical results on the force components from the attractive (\textbf{Left}) and repulsive (\textbf{Right}) components of the integrated interaction between a pair of thin rods in a skew configuration in 3-dimensions.}
    \label{fg:force_comparison}
\end{figure}


\begin{figure}[htb]
    \centering
    \includegraphics[width=1\textwidth]{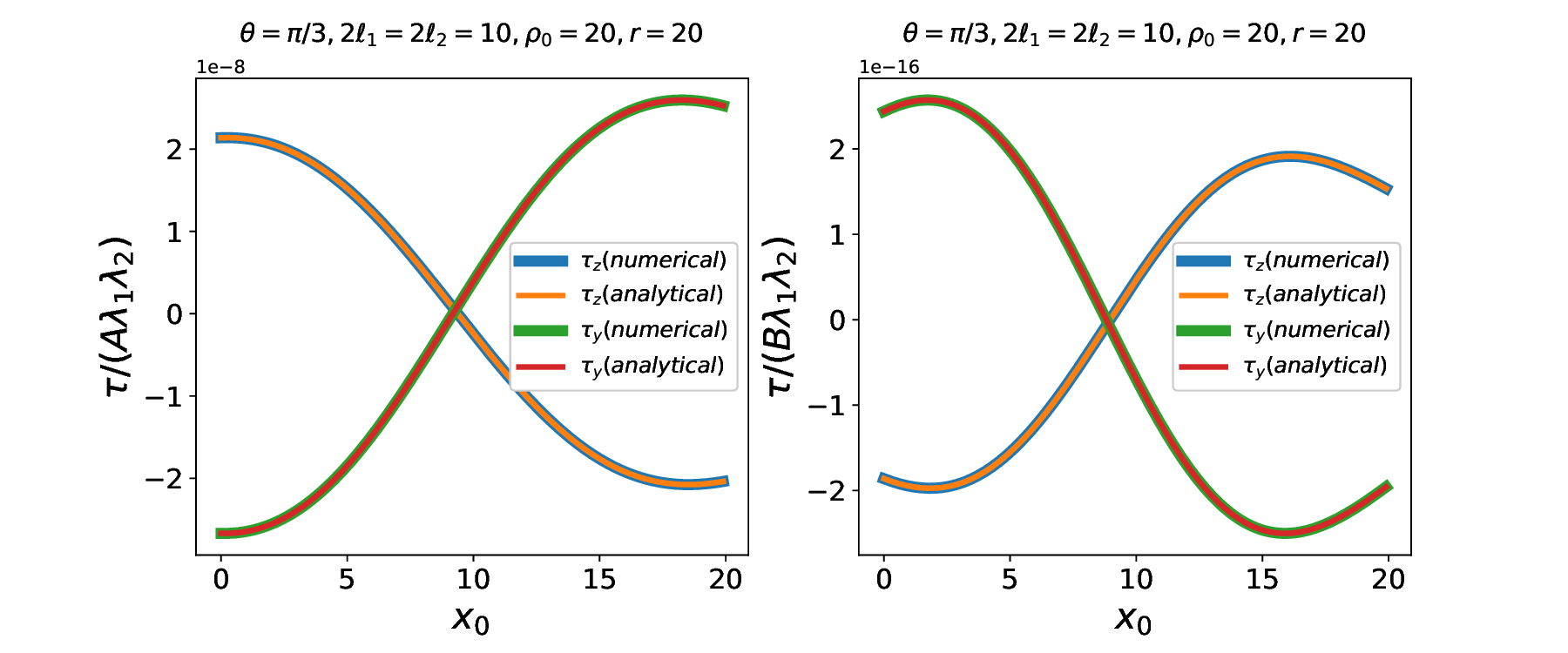}
    \caption{Comparison of analytical and numerical results on the torque components from the attractive (\textbf{Left}) and repulsive (\textbf{Right}) components of the integrated interaction between a pair of thin rods in a skew configuration in 3-dimensions.}
    \label{fg:torque_comparison}
\end{figure}


\begin{figure}[htb]
    \centering
    \includegraphics[width=1.0\textwidth]{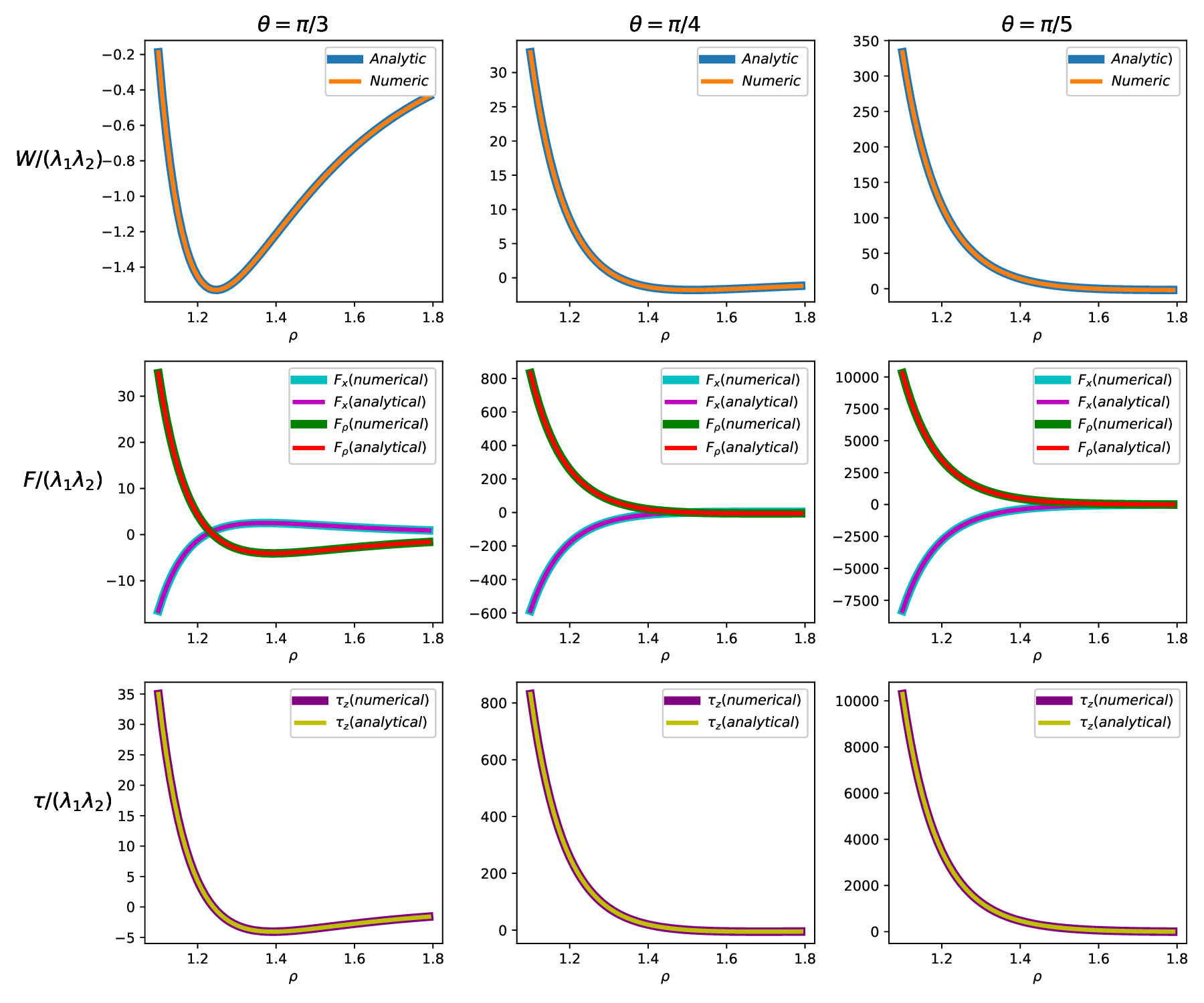}
    \caption{Comparison of analytical and numerical results for the integrated potentials, force components, and torque components between a bead and a thin rod with a length of $2l=10$ and $A=B=4$.}
    \label{fg:beadrod_comparison}
\end{figure}


\end{document}